\documentclass[twocolumn,showpacs,pre]{revtex4}

\usepackage{amsfonts,amssymb,amsmath,graphicx}
\newcommand{\ud}{\mathrm{d}}
\newcommand{\ic}{\mathrm{i}}
\newcommand{\im}{\mathrm{ Im }\ }

\begin{document}

\title{Trace formula for dielectric cavities II:\\ Regular,
  pseudo-integrable, and chaotic examples}
\author{E. Bogomolny$^{1,2}$, N. Djellali$^3$, R. Dubertrand$^4$,
  I. Gozhyk$^3$, M. Lebental$^{3}$,
  C. Schmit$^{1,2}$, C. Ulysse$^5$, and J. Zyss$^{3}$}
\affiliation{ $^{1}$Univ. Paris-Sud,
Laboratoire de Physique Th\'eorique et Mod\`eles statistiques,
Orsay, F-91405,\\
$^2$CNRS, UMR8626, Orsay, F-91405,\\
$^{3}$Ecole Normale Sup\'erieure de Cachan, CNRS, UMR 8537,
Laboratoire de Photonique Quantique et Mol\'eculaire, F-94235
Cachan,
France,\\
$^4$Institut f\"ur theoretische Physik, Philosophenweg 19
D-69120 Heidelberg, Germany,\\
$^5$CNRS, UPR20, Laboratoire de Photonique et Nanostructures, Route
de Nozay, F-91460 Marcoussis, France. }

\begin{abstract}
Dielectric resonators are open systems particularly interesting due
to their wide range of applications in optics and photonics. In a
recent paper [PRE \textbf{78}, 056202 (2008)] the trace formula for
both the smooth and the oscillating parts of the resonance density
was proposed and checked for the circular cavity. The present paper
deals with numerous  shapes which would be integrable (square,
rectangle, and ellipse), pseudo-integrable (pentagon) and chaotic
(stadium), if the cavities were closed (billiard case). A good
agreement is found between the theoretical predictions, the
numerical simulations, and experiments based on organic
micro-lasers.
\end{abstract}
\maketitle

\section{Introduction}

Open quantum (or wave) systems are rarely integrable and therefore
difficult to deal with. Over recent years, this field of research
has raised many crucial questions and  various systems have been
investigated. Here we consider open dielectric resonators for their
wide range of applications in optics and photonics
\cite{vahala,matsko}. In a first paper \cite{bogomolny}, the trace
formula for these systems was derived in the semi-classical regime
to infer their spectral features. More specifically in that paper
both the expressions for the weighting coefficients of the periodic
orbits and the counting function $N(k)$ (mean number of resonances with a
real part of the wave number less than $k$) were obtained and
demonstrated analytically for two integrable cases, the
two-dimensional (2D) circular cavity and the 1D Fabry-Perot
resonator. In the present paper, we consider in detail 2D dielectric
cavities with different shapes where no explicit exact solution is
known.  We compare the predictions of formulae obtained in
\cite{bogomolny} with numerical
simulations and experiments based on organic micro-lasers.

Resonance problems can be seen as counterparts of the
scattering of an electromagnetic wave on a finite obstacle.
This point of view turns out to be particularly interesting since
such scattering problems have been extensively studied (see e.g. \cite{laxphillips}).
Rigorous results for the scattering of a wave on convex obstacles with Dirichlet boundary
conditions were proved in \cite{petkovpopov}. Physical approach to these problems has been discussed in \cite{uzy}.
More recently some theorems were demonstrated in \cite{robert1,robert2}.  The general
structure of the resonance spectrum on a transparent smooth obstacle
was studied in \cite{popovvodev}.

This paper is focused on careful investigations of spectral
properties for 2D convex dielectric resonators, which are the open
counterparts of the so-called 'quantum billiards'. The outline of
the paper is the following. The  formulas obtained in
\cite{bogomolny} are recalled and the numerical and experimental
techniques are described in Sec. \ref{first}. Then different cavity
shapes are explored and their properties are compared with what is
known for billiards. The square, rectangle, and ellipse cases are
gathered in Sec. \ref{second}. We call such shapes  'regular shapes'
since the corresponding billiard problems are separable. In Sec.
\ref{third}, the pentagonal dielectric cavity was chosen to
illustrate a pseudo-integrable system. Eventually in Sec.
\ref{forth}, the Bunimovich stadium is investigated as an archetype
of a chaotic system. For completeness in
Appendix~\ref{sec:app-weyl}, the derivation of the Weyl's law is
briefly presented.

\section{Background: theory, numerics, and experiments}\label{first}

Real dielectric resonators  are  three-dimensional (3D) cavities
requiring that the 3D vectorial Maxwell equations are used. When the
cavity thickness is of the order of the wavelength, this problem can
be approximated to a 2D scalar equation following  the effective
index model, which is widely used in photonics (see e.g.
\cite{vahala} and references therein). This approach has been proved
to be quite efficient for our organic micro-lasers
\cite{lebental,lebental-matsko}. Briefly, it assumes that the
electromagnetic field can be separated in two independent
polarizations, called TM (resp. TE) if the magnetic (resp. electric)
field lies in the plane of the cavity ($xy$) \footnote{This
definition is consistent all over this paper. In the literature,
these names are sometimes permuted.}. In this 2D approximation the
Maxwell equations are reduced to the Helmholtz equation
\begin{equation}
(\Delta_{xy}+n^2k^2)\,\psi=0
\label{eq:onde-base}
\end{equation}
where $\psi$ stands for the $z$-component of the electric (resp.
magnetic) field in TM (resp. TE) polarization. After resolution, all
the components of the electromagnetic fields can be inferred from
$\psi$. In Eq. (\ref{eq:onde-base}) $k$ is the wave number and $n$
the effective refractive index. It is worth highlighting that the
error of this approximation is not well controlled \cite{bittner_2}.

The boundary conditions in this 2D approximation are the following:
\begin{eqnarray}
\psi_1=\psi_2,\hspace{0.5cm}&\mbox{and}&\hspace{0.9cm}
\frac{\partial \psi_1}{\partial \nu}=\frac{\partial \psi_2}{\partial
  \nu}\hspace{1cm}\mbox{(TM)}\nonumber\\
\psi_1=\psi_2,\hspace{0.5cm}&\mbox{and}&\hspace{0.5cm}
\frac{1}{n^2}\frac{\partial \psi_1}{\partial \nu}=\frac{\partial
\psi_2}{\partial \nu} \hspace{1cm}\mbox{(TE)} \nonumber
\end{eqnarray}
where $\nu$ is a direction normal to the boundary and $\psi_1$
(resp. $\psi_2$) corresponds to the field inside (resp. outside) the
cavity. In the case of an open system such as a dielectric cavity,
the resonances are defined as the solutions of (\ref{eq:onde-base})
with the outgoing boundary condition at infinity:
\begin{equation}
  \psi(\vec{x})\propto e^{\ic k |\vec{x}|}\quad  |\vec{x}|\to \infty
\label{eq:BCinfinity}
\end{equation}
Then the resonance eigenvalues, $k_n^2$, are complex with negative
imaginary part
\begin{equation}
  \label{eq:def-E-tau}
  k_n^2=E-\frac{\ic}{2\tau}\ .
\end{equation}
$E$ is called the energy of the resonance whereas $\tau$ is its
lifetime. The wave numbers of the low-loss resonances (higher
quality factors) are thus located closed to the real axis.

\subsection{Semiclassical trace formula}

Here for simplicity, we consider only TM polarization where the
functions and their normal derivative are continuous on the
cavity boundary. In this case, it appears that the resonance spectrum
splits into two subsets, depending on the imaginary part of the wave
numbers. For one of the subsets, the wave numbers lie above a
boundary
\begin{equation}
  \label{inner}
  \gamma_{max}< \im k_n  < 0\ ,
\end{equation}
where $\gamma_{max}$ is a certain constant which depends on the
cavity, and the corresponding wave functions are mainly concentrated
inside the cavity. These resonances are similar to the so-called
Feschbach resonances. For the second class of resonances (called
shape resonances) the wave functions are mainly supported outside
the cavity and the corresponding eigenvalues have large imaginary
parts. For smooth convex obstacles, it was shown (see e.g.
\cite{sjostrand} and references therein) that they obey the
inequality
$$\textrm{Im}\, k_n < -\,\textrm{Const}\, |\textrm{Re}\, k_n|^{1/3}. $$
Hereafter we will focus only on Feschbach (inner) resonances, since
they are the most relevant for lasers and photonics applications.
They will simply be referred to as ``resonances'' from now on.

The spectral density can formally be separated in two contributions
\begin{equation}
  \label{eq:densite-generale}
  d(k)=\overline{d}(k)+d^{(osc)}(k)\ .
\end{equation}
$\overline{d}(k)$ stands for the smooth part and is usually written
through the counting function $\overline{d}(k)=\ud \bar{N}(k)/\ud k$
which counts how many resonances in average have a real part less
than $k$ \footnote{When computing $N(k)$ numerically we did not use
any averaging, so we just wrote $N(k)$.}. The oscillating part,
$d^{(osc)}$, can be related in the semiclassical regime $kl\gg 1$
($l$ is any characteristic length of the cavity) to a sum over the
classical periodic orbits \cite{gutzwiller}.

In \cite{bogomolny}, the semiclassical trace formula for open
dielectric cavities was derived. It states that the counting
function of dielectric resonators can be written as follows:
\begin{equation}
\label{eq:smooth}
  \bar{N}(k)=n^2 \displaystyle\frac{{\cal A} k^2}{4\pi}+\tilde{r}(n)
  \displaystyle\frac{{\cal L}k}{4\pi}+ {\cal O}(1)
\end{equation}
where ${\cal A}$ is the area of the cavity, ${\cal L}$ its
perimeter, and $\tilde{r}(n)$ a function of the refractive index
involving elliptic integrals:
$$\tilde{r}(n)=1+\frac{n^2}{\pi}\int_{-\infty}^{\infty}\frac{\ud
t}{t^2+n^2}\, R(t)-\frac{1}{\pi}\int_{-\infty}^{\infty}\frac{\ud
t}{t^2+1}\, R(t)$$ with
$$R(t)=\frac{\sqrt{t^2+n^2}-\sqrt{t^2+1}}{\sqrt{t^2+n^2}+\sqrt{t^2+1}}$$
The derivation of (\ref{eq:smooth}) and details on $\tilde{r}(n)$
and $R(t)$ are given in Appendix \ref{sec:app-weyl}. In the
following, we will compare for various shapes the prediction of
(\ref{eq:smooth}) to the function $N(k)$ inferred from numerical
simulations, and show a good agreement in all considered cases. In
particular, we will stress the non trivial linear coefficient
\begin{equation}\label{eq:smooth-lin}
\alpha^{th}=\tilde{r}(n)
  \displaystyle{\cal L}/4\pi
\end{equation}
In this paper, in general $n=1.5$, and so $\tilde{r}(1.5)=1.025$.

The oscillating part of the trace formula is written as a sum over
classical periodic orbits ($p.o.$):
\begin{equation}
\label{eq:osc} d^{(osc)}(k)= \sum_{p.o.} \big(c_p \,e^{\ic n k l_p}+
c.c. \big)
\end{equation}
where $l_p$ is the length of the orbit and $c_p$ its amplitude which
depends only on classical quantities. We count a periodic orbit and
the corresponding time-reversed orbit as a single orbit. The
expressions for the $c_p$ can be derived in a standard way (see eg.
\cite{gutzwiller,bb}), using the formula of the reflected Green
function given in Appendix \ref{sec:app-weyl}. As for billiard, it
depends whether the orbit is isolated (i.e. unstable) or not. For an
isolated periodic orbit
\begin{equation}
  \label{eq:cp-isolee}
 c_p=\frac{2nl_p}{\pi}\frac{1}{|\det(M_p-1)|^{1/2}} R_p \,e^{-\ic
      \mu_p\pi/2}
\end{equation}
where $M_p$, $\mu_p$, and $R_p$ are respectively the monodromy
matrix, the Maslov index of the orbit, and the product of the
Fresnel reflection coefficients at all reflection points. For a ray
with an angle of incidence $\chi$, the TM Fresnel reflection
coefficient at a planar dielectric interface between a medium with a
refractive index $n$ and air is:
\begin{equation}
  \label{eq:fresnel}
     R_{TM}(\chi)=\frac{n\cos\chi-\sqrt{1-n^2\sin^2\chi}}{n\cos\chi +
    \sqrt{1-n^2\sin^2\chi}}
\end{equation}
For a periodic orbit family
\begin{equation}
  c_p=\frac{\sqrt{2k}}{\pi}\frac{n^2\mathcal{A}_p}{\sqrt{\pi nl_p}}
  \left<R_p\right> e^{ -\ic \mu_p\pi/2+\pi/4}\ , \\
 \label{eq:cp-famille}
\end{equation}
where $\mathcal{A}_p$ is the area covered by the orbit family and
$\left<R_p\right>$ stands for the average of the Fresnel reflection
coefficient over the family.

Hereafter, to compare these theoretical predictions to numerical
simulations, we will rather consider the Fourier transform of the
spectral density $d(l)$ in order to reveal the oscillating part:
\begin{equation}
  \label{eq:d-l}
     d(l)=\sum_n e^{-ik_nl}
\end{equation}
where the $k_n$ are the complex eigenvalues calculated from
numerical simulations. This function can be obtained from
experiments as well.

\subsection{Numerical simulations}

The numerical simulations are based on the Boundary Element Methods
which consists of writing the solution of (\ref{eq:onde-base}) as
integral equations on the inner and outer sides of the boundary and
of matching them using the boundary conditions. The complex spectrum
and the resonance wave functions (sometimes called quasi-stationnary
states or quasi-bound states) are inferred from the obtained
boundary integral equation. In accordance with the experiments
presented here based on polymer cavities, we used $n=1.50$ inside
the cavity and $n=1$ outside (air).

\subsection{Experiments}

\begin{figure}
\begin{minipage}[l]{.45\linewidth}
\includegraphics[width=.5\textwidth]{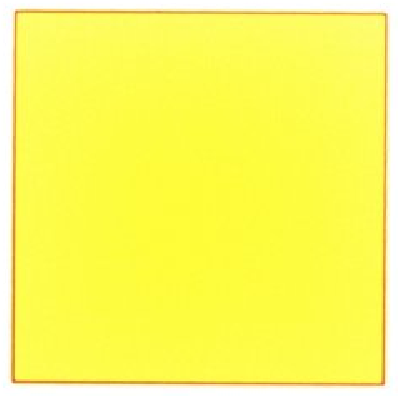}
\end{minipage}
\begin{minipage}[l]{.45\linewidth}
\includegraphics[width=.6\textwidth]{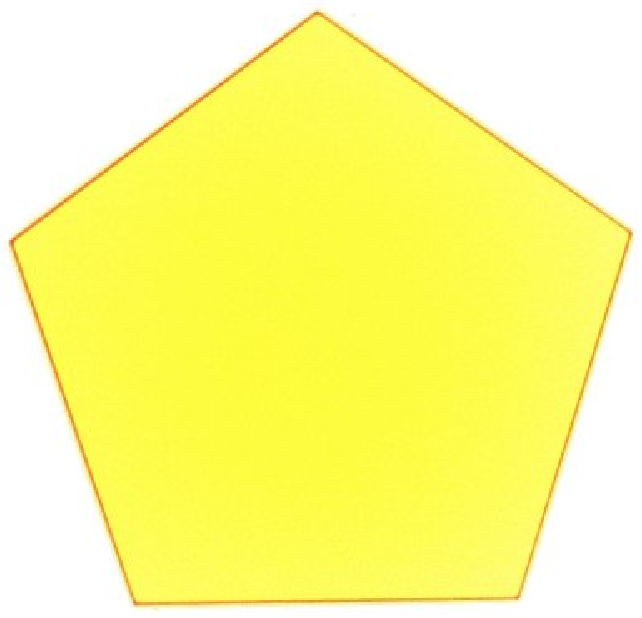}
\end{minipage}
\begin{minipage}[l]{.45\linewidth}
\includegraphics[width=.7\textwidth]{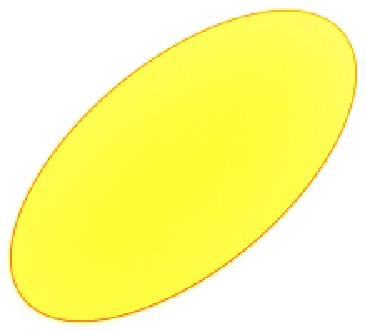}
\end{minipage}
\begin{minipage}[l]{.45\linewidth}
\includegraphics[width=.8\textwidth]{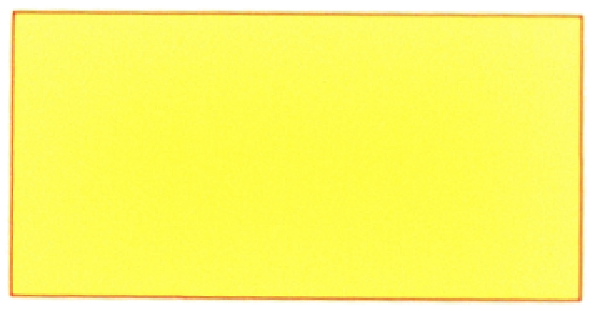}
\end{minipage}
\begin{minipage}[r]{.45\linewidth}
\includegraphics[width=.7\textwidth]{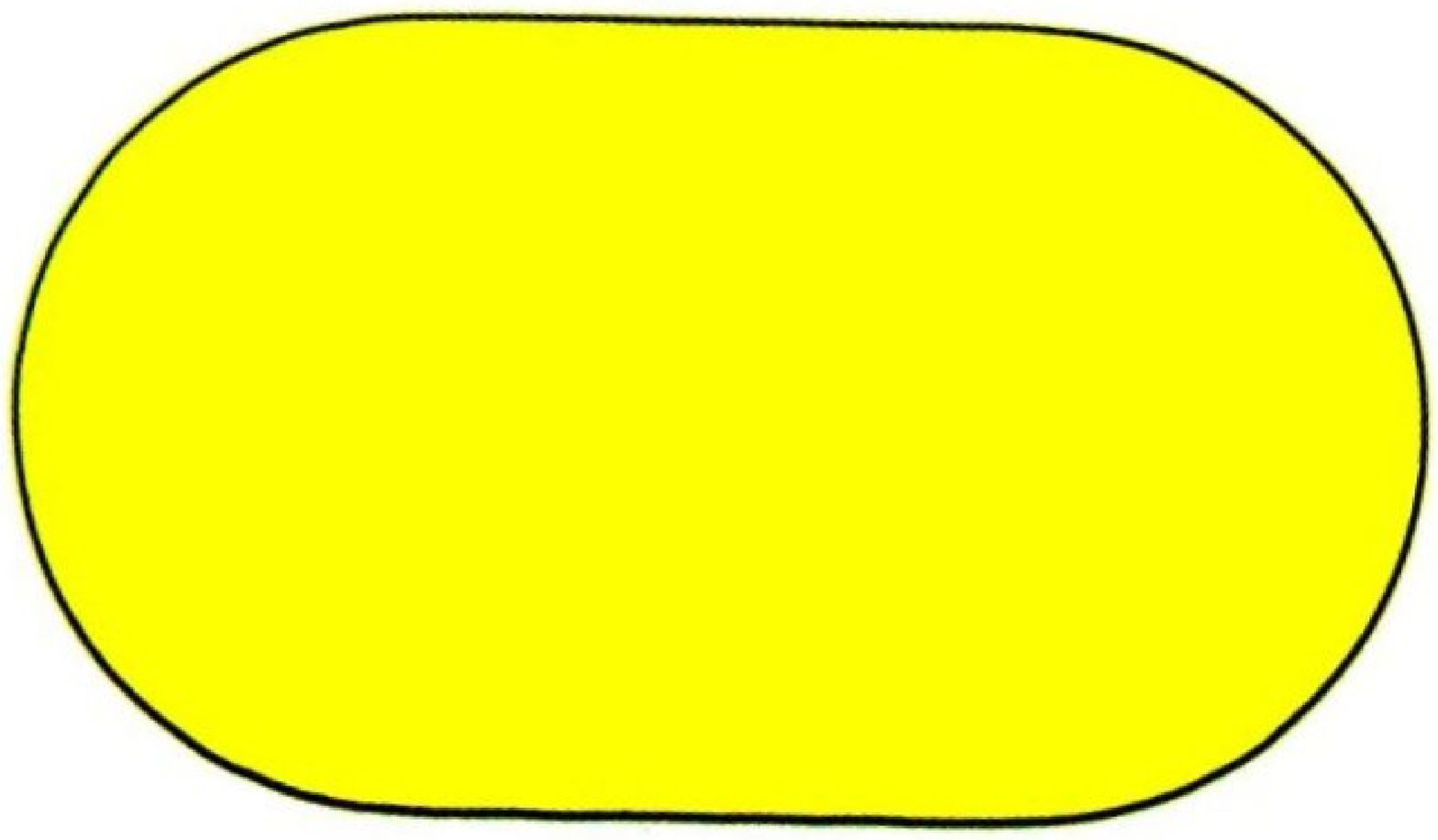}
\end{minipage}
\caption{(Color online) Photographs through an optical microscope of
some micro-lasers used in this paper. The in-plane scale is about
100 $\mu$m.} \label{fig:photos}
\end{figure}

Dielectric resonators are widely used in photonics for fundamental
research \cite{chang} and practical applications \cite{matsko}.
Furthermore their wavelength range is not limited to optics and
cover other electromagnetic domains like microwaves
\cite{micro-ondes} or Tera-Hertz waves \cite{preu}. Here we consider
quasi-2D organic micro-lasers since they proved to be quite
efficient to test trace formulae \cite{lebental,lebental-matsko}.

The cavities were etched by electron-beam lithography into a
polymethylmethacrylate (PMMA) layer doped with a laser dye
\footnote{4-dicyanomethylene-2-methyl-6-(4-dimethylaminostyryl)-4H-pyran,
(DCM), 5 \% in weight.} spincasted on a silica on silicon wafer.
This technology offers appreciable versatility in terms of shapes,
while ensuring a small roughness and a good quality for the corners
with a resolution better than a tenth of wavelength (see Fig. 6.5 in
\cite{lebental-matsko}). Some photographs of the cavities studied in
this paper are shown in Fig. \ref{fig:photos}. The fabrication
process is relatively fast and reproducible. At the end, the cavity
thickness is about 600 nm, while the in-plane scale is of the order
of a few dozens of microns, which allows to apply the effective
index approximation and
therefore to consider these cavities as 2D resonators.

The chosen cavity was uniformly pumped from above at room
temperature and atmosphere with a frequency-double pulsed Nd:YAG
laser (30 or 700 ps) and its emission, integrated over 30 pump
pulses, was collected sideways in its plane with a spectrometer
(Acton SpectraPro 2500i) coupled to a cooled CCD camera (Pixis100
Princeton Instruments). The spectral range of the emitted light
depends on the dye laser. Here, for DCM, it is centered around 600
nm and so the $kl$ parameter varies from 500 to 1000.
Consequently this experimental system is working far away within the
semi-classical regime while its coherence properties are ensured by
lasing.

The laser emission is mostly TE polarised \cite{gozhyk}, but for the
features which are compared here with theory and numerics, there is
not any predicted difference between the TE and TM cases. Among the
resonator shapes studied in this paper, the pump polarisation plays
a prominent role only for the square where it will be further
developed. For the other shapes, it will not be mentioned.

As this paper is focusing on spectral features, we will consider
only the emission spectra which, by default, were registered in the
direction of maximal emission (i.e. parallel to the sides for square
and pentagon \cite{djellali}, parallel to the shortest axis for the
rectangle \cite{djellali}, and at an angle depending on the shape
parameter for stadiums \cite{lebental2}). Moreover in order to be
closed enough to the theoretical case of a passive resonator, the
cavities were pumped just above the laser threshold. Mode (and
orbit) competition is then reduced.

The typical laser spectrum is made of one or several combs of peaks
connected (in a crude approximation)  with  certain periodic orbits. As shown in
\cite{lebental} the geometrical lengths of the underlying periodic
orbits can be inferred from the Fourier transform of the
experimental spectrum, which is an equivalent of the length density
$d(l)$. For instance, for a Fabry-Perot resonator of width $a$, the
geometrical length of the single periodic orbit is $L=2a$ and the
dephasing after a loop should be a multiple of $2\pi$: $knL=2\pi m$
with $m\in\mathbb{N}^{\star}$. Then the spacing between the comb
peaks verifies $\Delta k = 2\pi /nL$, leading to a periodic comb
pattern and a Fourier transform of the spectrum peaking at $nL$.
With our experimental set-up, the precision on the geometrical
length $L$ reaches 3 $\%$ after taking duly account of the
dispersion due to the effective index and the absorption of the
laser dye. So the refractive index which should be used to interpret
the Fourier transform is 1.64 for these actual experiments (it is
different from the bulk refractive index 1.54 and the effective
refractive index 1.50) \cite{lebental}.

\section{Regular shapes}\label{second}

This section deals with square, rectangle, and elliptic dielectric
cavities, which can be called 'regular' cases since their closed
counterparts (billiard problems) are integrable. To our knowledge no
analytical solution has been proposed so far for these cavity shapes in the
open case. Nevertheless, their dielectric spectrum  shows some
characteristic features specific to integrable systems.

The only example of 2D integrable dielectric cavity is the circular
one (see e.g. \cite{dubertrand}).  Therefore its resonances are
organized in regular branches labeled by well defined quantum
numbers. For the above mentioned 'regular'  cavities  with
relatively small refractive index it appears  that the resonances
still follow similar branch structure (see Figs.~\ref{fig:carre}, \ref{fig:carre_n1.35}, \ref{specrecta2}, \ref{specell}). This is surprising as these
dielectric  problems are not integrable and strictly speaking there
is no conserved quantum number. This unusual regularity can be
described by the superscar approximation proposed in
\cite{lebental}. The detailed discussion of the modified superscar
model and its application to these problems will be given elsewhere
\cite{Eugene_square}.

\subsection{The square cavity}

The simplest example of regular cavities is the dielectric square
where the inside billiard problem is straightforward. For instance,
for Dirichlet boundary conditions, the eigenenergies and
eigenfunctions are the following:
\begin{eqnarray*}
  k_{p,m}^2&=&\frac{\pi^2}{a^2}(p^2+m^2)\ ,\\
  \psi_{p,m}(x,y)&=&\frac{2}{a}\sin\left(
    \frac{p\pi x}{a} \right)\sin\left(
    \frac{m\pi y}{a} \right)\ ,
\end{eqnarray*}
where $a$ is the side length and $p$ and $m$ are two positive
integers. The outer scattering problem even with the Dirichlet
boundary conditions is more difficult as it corresponds to a
pseudo-integrable problem (see e.g. \cite{pseudoint}) and no
explicit analytical solution exists \cite{barbara}.

\subsubsection{Numerics}

\begin{figure}
\begin{minipage}[l]{.45\linewidth}
\includegraphics[width=.8\textwidth]{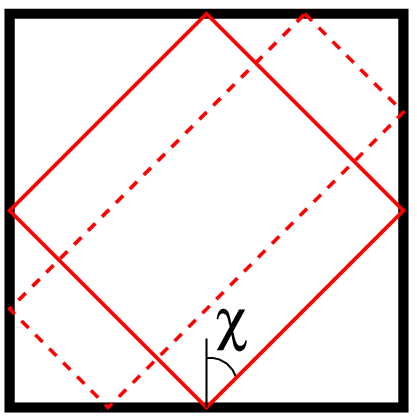}
\centering (a)
\end{minipage}
\begin{minipage}[r]{.45\linewidth}
\includegraphics[width=.8\textwidth]{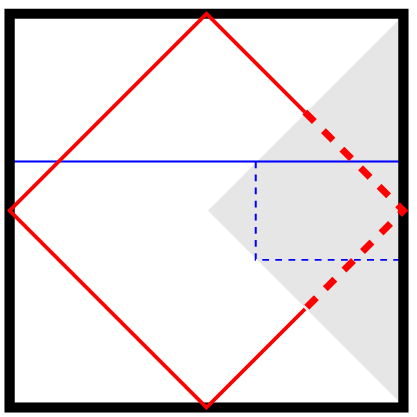}
\centering (b)
\end{minipage}
\caption{(a) Two representations of the same diamond periodic orbit.
Definition of the incident angle $\chi$. (b) The dashed part
corresponds to the fundamental domain used for numerical
simulations. The Fabry-Perot and diamond periodic orbits are drawn,
as well as their restriction to the fundamental domain (dotted
lines).} \label{fig:schema-carre}
\end{figure}

\begin{figure}
\includegraphics[width=1\linewidth]{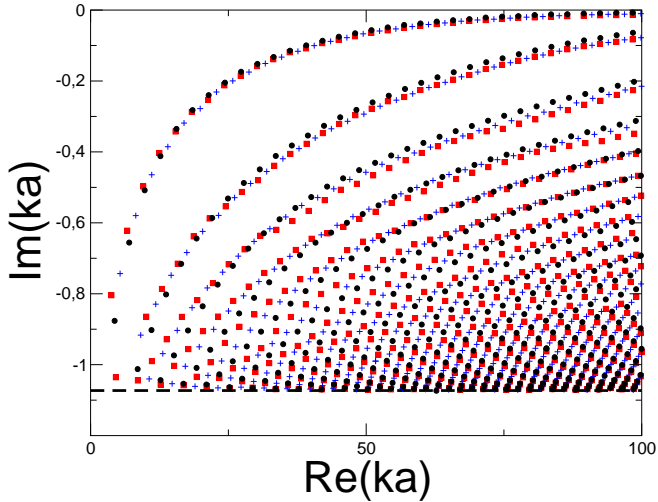}
\caption{(Color online) Resonance spectrum for the dielectric square
with $n=1.5$. The crosses, squares, and circles correspond
respectively to $(-+)$, $(++)$, and $(--)$ symmetry classes. The
position of the horizontal dashed line is given by (\ref{bbsq}).}
    \label{fig:carre}
\end{figure}

\begin{figure}
  \begin{minipage}[l]{.98\linewidth}
\includegraphics[width=.5\linewidth]{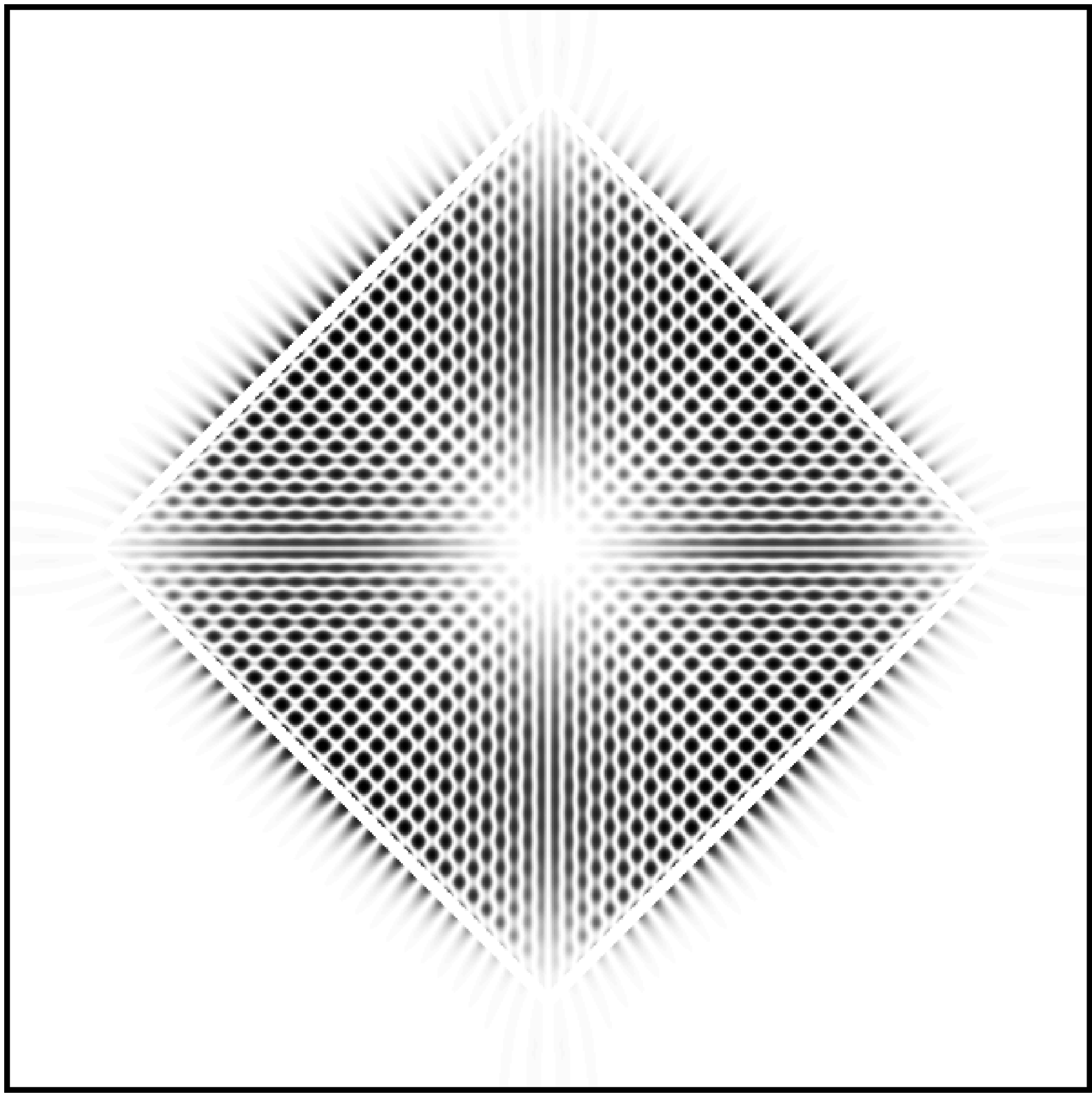}
    { \centering (a)}
  \end{minipage}
  \begin{minipage}[r]{1\linewidth}
    \includegraphics[width=.5\linewidth]{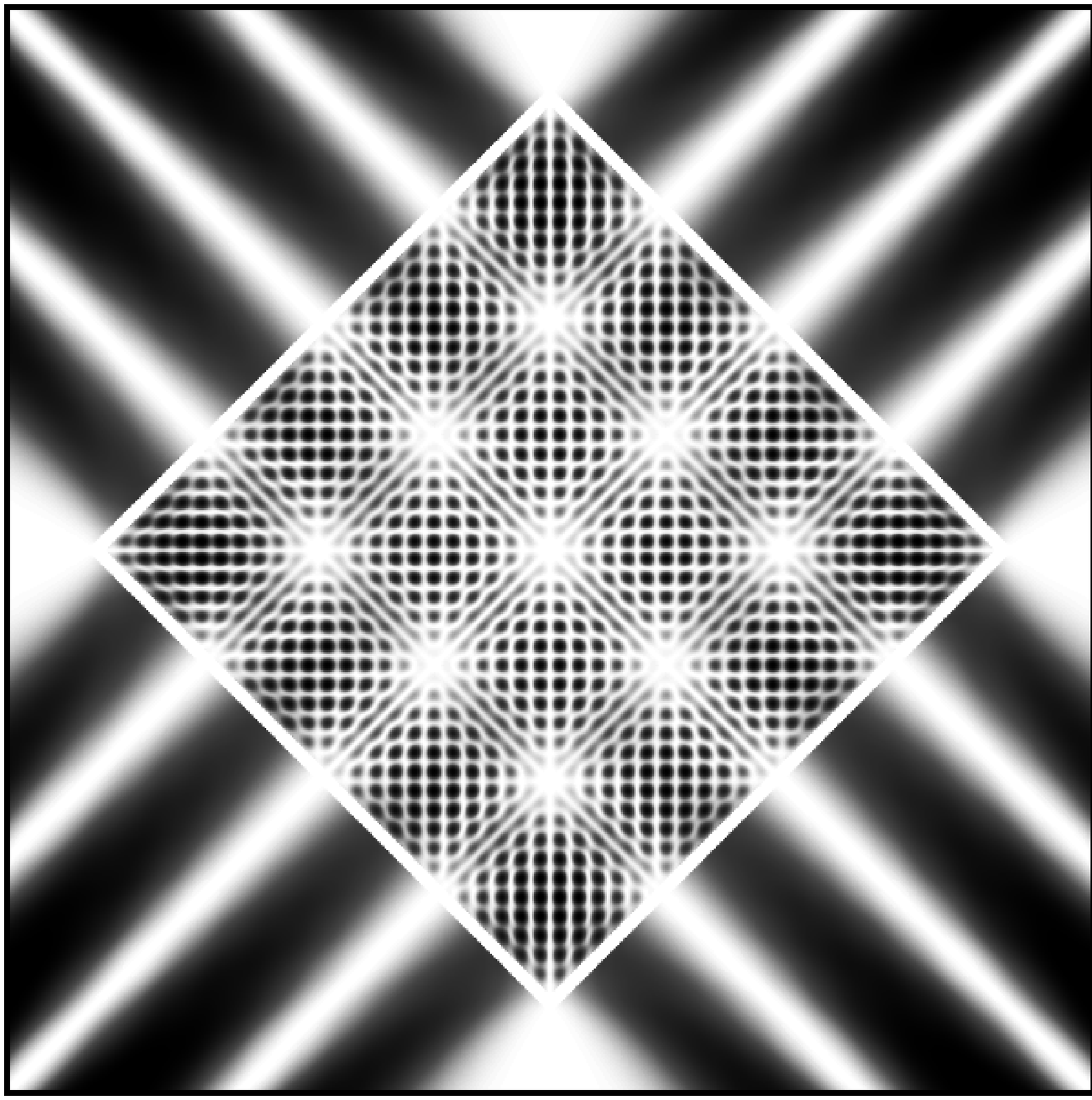}
    { \centering (b)}
  \end{minipage}
\begin{minipage}[r]{1\linewidth}
    \includegraphics[width=.5\linewidth]{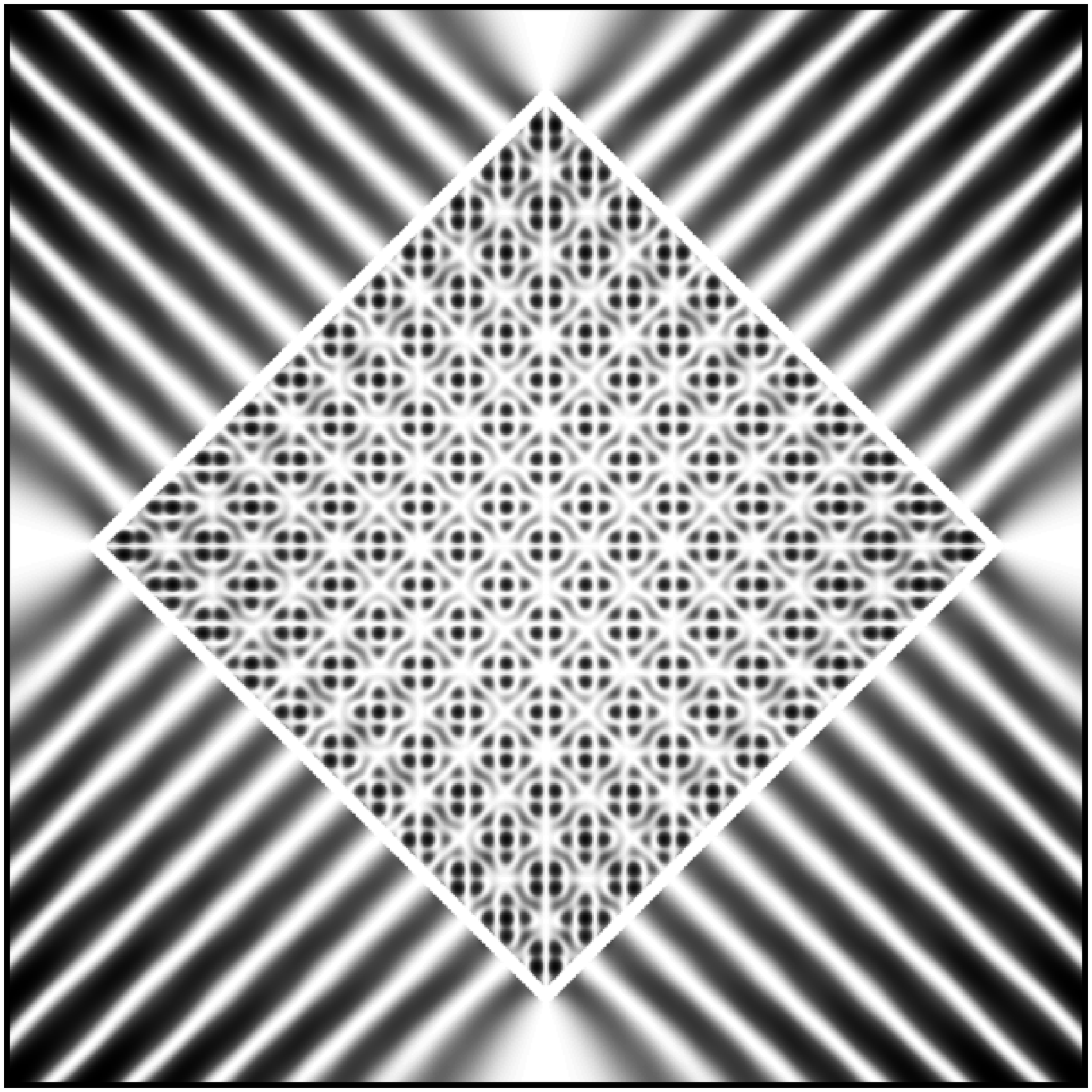}
    { \centering (c)}
  \end{minipage}
\caption{Wave functions for the dielectric square with $n=1.5$ and
$(- -)$ symmetry class. (a), (b), and (c) correspond to $ka=98.36
-0.0071\,\ic$, $ka=98.78-1.061\,\ic$, and $ka=98.96-0.998\,\ic$.
Greyscale, black representing maximal values of $|\psi|^2$.}
\label{wave_carre}
\end{figure}

The solutions of the square dielectric problem can be divided into
four symmetry classes corresponding to wave functions  odd $(-)$ or
even $(+)$ with respect to the diagonals $y=x$ and $y=-x$. For
instance, the notation $(-+)$ means that the wave function is odd
with respect to the diagonal $y=x$ and even with respect to the
other. Each symmetry class  reduces to a quarter of a square (dashed
part in Fig.~\ref{fig:schema-carre}b) with Dirichlet, $(-)$, or
Neumann, $(+)$, boundary conditions along the diagonals. The $(-+)$
and $(+-)$ symmetry classes are equivalent. Fig. \ref{fig:carre}
shows the resonance spectrum for all symmetry classes and $n=1.5$.
Fig.~\ref{wave_carre} displays some typical quasi-stationary states
for the $(- -)$ symmetry class from different parts of the spectrum.
Notice highly unusual regularity of the spectrum and wave functions
for this shape.

As for the  case of the circular cavity (see e.g.
\cite{dubertrand}), the imaginary part of the dielectric square
resonances is bounded by the losses $\gamma_{max}$ of the periodic orbit
with the highest losses, which is here the Fabry-Perot indicated in Fig.~\ref{fig:schema-carre}b with
\begin{equation}
  \label{bbsq}
 \gamma_{max}=\frac{1}{n}\ln\left(\frac{n-1}{n+1}\right)\Big |_{n=1.5}\simeq -1.073
\end{equation}
The counting function $N(k)$ gives the mean number of resonances with a
real part less than $k$ in the strip defined by (\ref{inner}). The
Weyl-type formula (\ref{eq:smooth}) estimates its growth when $k\to
\infty$. We checked this prediction for different values of the
refractive index and each symmetry class. For $n=1.5$ the results of
the numerical fit to the data  computed from the spectrum in
Fig.~\ref{fig:carre} are the following
\begin{eqnarray*}
  (--)\,\, N_{fit}(k)&=&\frac{n^2}{16\pi}(ka)^2-0.0866 \,ka-0.0456,\\
  (++)\,\, N_{fit}(k)&=&\frac{n^2}{16\pi}(ka)^2+0.2489 \,ka-1.737,\\
  (-+)\,\, N_{fit}(k)&=&\frac{n^2}{16\pi}(ka)^2+0.0806 \,ka-2.012.
\end{eqnarray*}
Here we fixed the coefficient of the quadratic term and fitted the
linear and constant terms from the numerical data.

\begin{figure}
\includegraphics[width=0.8\linewidth]{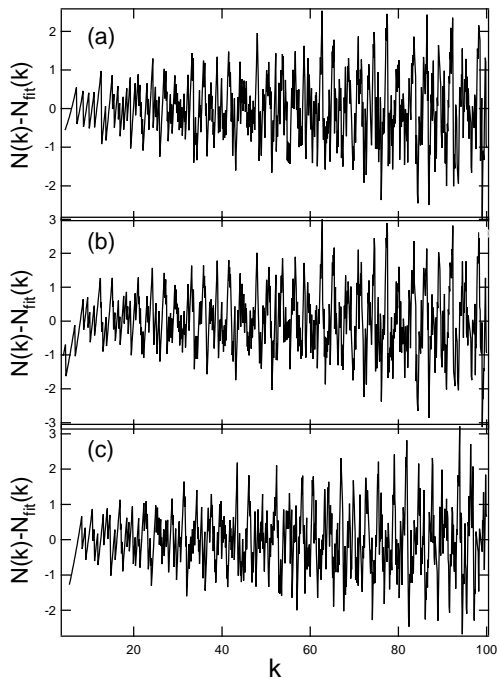}
\caption{Residue $N(k)-N_{fit}(k)$ for the dielectric square ($a=1$)
with  $n=1.5$ and different symmetry classes : (a) $(--)$, (b)
$(++)$, and (c) $(-+)$. }
\label{fig:N_k_carre}
\end{figure}

\begin{figure}
\includegraphics[width=0.9\linewidth]{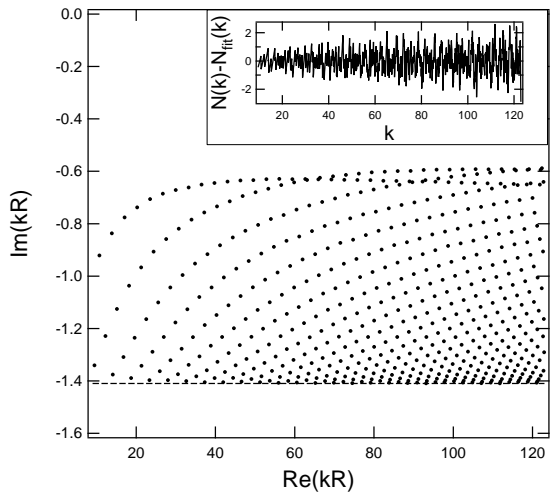}
\caption{Resonance spectrum for the dielectric square with
  $n=1.35$ and $(--)$ symmetry class. The positions of the horizontal
  dashed lines are given by (\ref{bbsq}). Insert: $N(k)-N_{fit}(k)$ with $a=1$.}
\label{fig:carre_n1.35}
\end{figure}

The predictions of (\ref{eq:smooth-lin}) which take into account the
Dirichlet or Neumann boundary conditions \footnote{$\bar{N}(k)=
\displaystyle\frac{{\cal A} k^2}{4\pi}+r
  \displaystyle\frac{{\cal L}k}{4\pi}+ {\cal O}(1)$ with $r=\pm 1$ for billiards
  with, respectively, Neumann and Dirichlet
  boundary conditions.} on two of the sides
of the fundamental domain are  given by the following expressions
calculated for $n=1.5$
\begin{eqnarray*}
  (--)\,\, \alpha^{th}=&\dfrac{\tilde{r}(n)-n\sqrt{2}}{4\pi}\Big|_{n=1.5}
  &\simeq -0.0872,\\
  (++)\,\,
  \alpha^{th}=&\dfrac{\tilde{r}(n)+n\sqrt{2}}{4\pi}\Big|_{n=1.5}&\simeq
  0.2504,\\
  (-+)\,\, \alpha^{th}=&\dfrac{\tilde{r}(n)}{4\pi}\Big|_{n=1.5} &\simeq
  0.0816. \\
\end{eqnarray*}
The predicted values are in good agreement with our numerical
calculations. More precisely, each residue (the difference between
$N(k)$ and $N_{fit}(k)$) oscillates around zero as evidenced in
Fig.~\ref{fig:N_k_carre}.
To ensure the efficiency of formula (\ref{eq:smooth}) we calculated
the spectrum also for $n=1.35$ with $(- -)$ symmetry class, see
Fig~\ref{fig:carre_n1.35}. The quadratic fit for the counting
function gives now
\begin{equation*}
  N_{fit}(k)=\frac{n^2}{16\pi}(ka)^2-0.0709\,ka-2.052
\end{equation*}
to be compared with:
\begin{equation*}
  \alpha^{th}=\dfrac{\tilde{r}(n)-n\sqrt{2}}{4\pi}\Big |_{n=1.35}\simeq -0.0713.
\end{equation*}
The residue stays also close to zero (see insert in
Fig.~\ref{fig:carre_n1.35}).\\
In all investigated examples, the agreement between prediction and
numerics is better than 2\% for the linear term.

The oscillatory part of the trace formula is checked as well. In the
square, periodic orbits form families. Thus their weighting is
predicted by (\ref{eq:cp-famille}), which implies that the spectrum
is dominated by the diamond periodic orbit (see
Fig.~\ref{fig:schema-carre}a). Actually the weighting coefficient of
this orbit is calculated as follows: it covers the whole cavity
(${\cal A}_\textrm{diamond}=a^2$), its length is short (${\cal
L}_\textrm{diamond}=2a\sqrt{2}$), and for $n>\sqrt{2}$ there is no
refractive loss ($|R_{diamond}|=1$) in the deep semiclassical limit
$k\to \infty$. For illustration, it is worth comparing with the
$c_p$ coefficient of the Fabry-Perot periodic orbit \footnote{In the
square there exist two identical Fabry-Perot orbits, horizontal and
vertical, each is self-retracing and so each is weighted by the
additional factor $1/2$.}: ${\cal A}_\mathrm{FP}=a^2$, ${\cal
L}_\mathrm{FP}=2a$, and $R_{\mathrm{FP}}=[(n-1)/(n+1)]^2$. Then for
$n=1.5$, $|c_{FP}/c_{diamond}|\simeq 0.05$ mainly due to the
prominent influence of refractive losses.

\begin{figure}
\centering\includegraphics[width=1\linewidth]{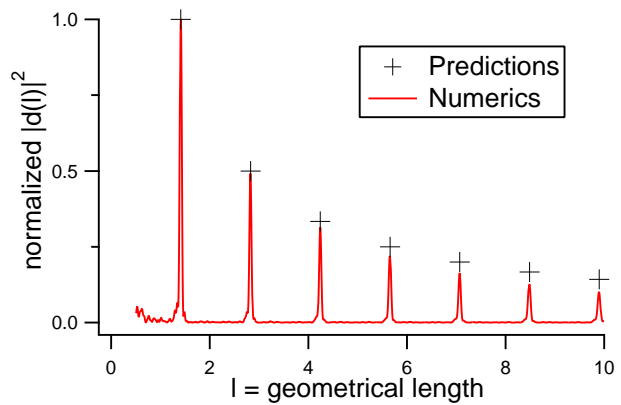}
\caption{Comparison of $|d(l)|^2$ calculated from the numerical data
of the dielectric square for $(--)$ symmetry class and $n=1.5$
(continuous line), and predictions from trace formula (crosses)
considering the $m^{th}$ repetitions of half the diamond periodic
orbit ($l=ma\sqrt{2}$) with a coefficient $|d(l)|^2\propto 1/m$.}
\label{fig:d-l-carre}
\end{figure}

The agreement with numerical simulations is checked via the length
density of the dielectric square which is computed from the
numerical spectrum using (\ref{eq:d-l}). The results are plotted in
Fig.~\ref{fig:d-l-carre}. The length density is highly peaked at
$l=a\sqrt{2}$ and at its harmonics ($l_m=ma\sqrt{2}$, with
$m\in\mathbb{N}^{\star}$). Only half of the diamond orbit length
appears, since the length density is calculated for a single
symmetry class and Fig.~\ref{fig:schema-carre}b shows that the
diamond periodic orbit is twice shorter if restricted to the dashed
area. If the length density had been performed with the four
symmetry classes, it would have been peaked at the full diamond
length ($2a\sqrt{2}$) and at its harmonics. The same appears with
experiments as shown below.

The agreement between numerics and predictions from trace formula
(\ref{eq:cp-famille}) is quite good as well when comparing the
ratios of the harmonics. Actually these harmonics can be identified
as repetitions of the diamond periodic orbit (${\cal L}=m{\cal
L}_{\textrm{diamond}}$) and thus formula (\ref{eq:cp-famille})
predicts that the $|c_p|^2$ should decrease like $1/m$. This
prediction is shown by crosses in Fig.~\ref{fig:d-l-carre}. From
numerics, we received harmonics a little bit smaller than predicted
which is natural as the Fresnel reflection coefficient
(\ref{eq:fresnel}) does not take into account correctly a leakage
through a dielectric interface of finite length.

\subsubsection{Experiments}

\begin{figure}
\begin{minipage}[t!]{1\linewidth}
\includegraphics[width=1\linewidth]{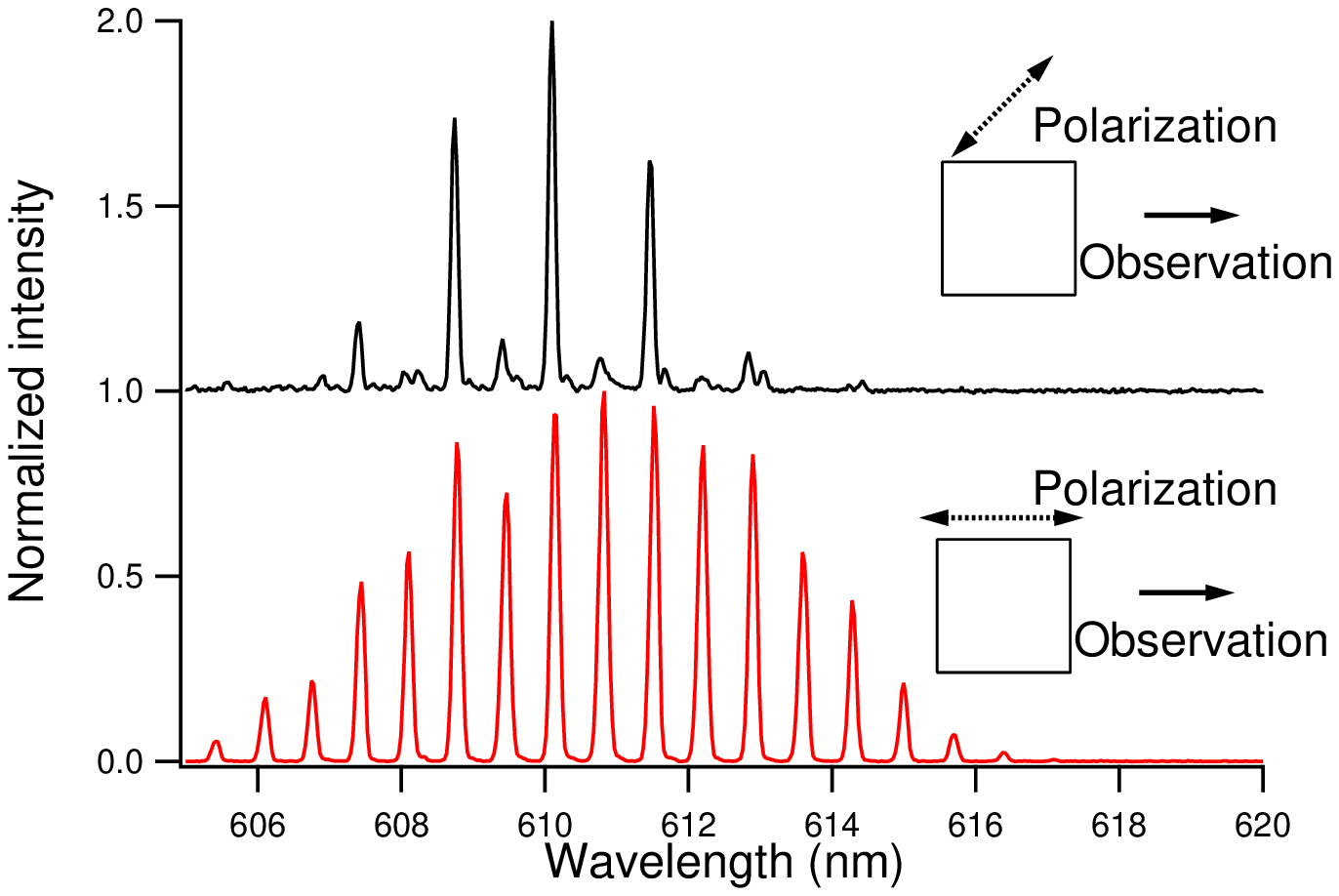}
\begin{center}(a)\end{center}
\end{minipage}\hfill
\begin{minipage}[t!]{1\linewidth}
\includegraphics[width=.8\linewidth]{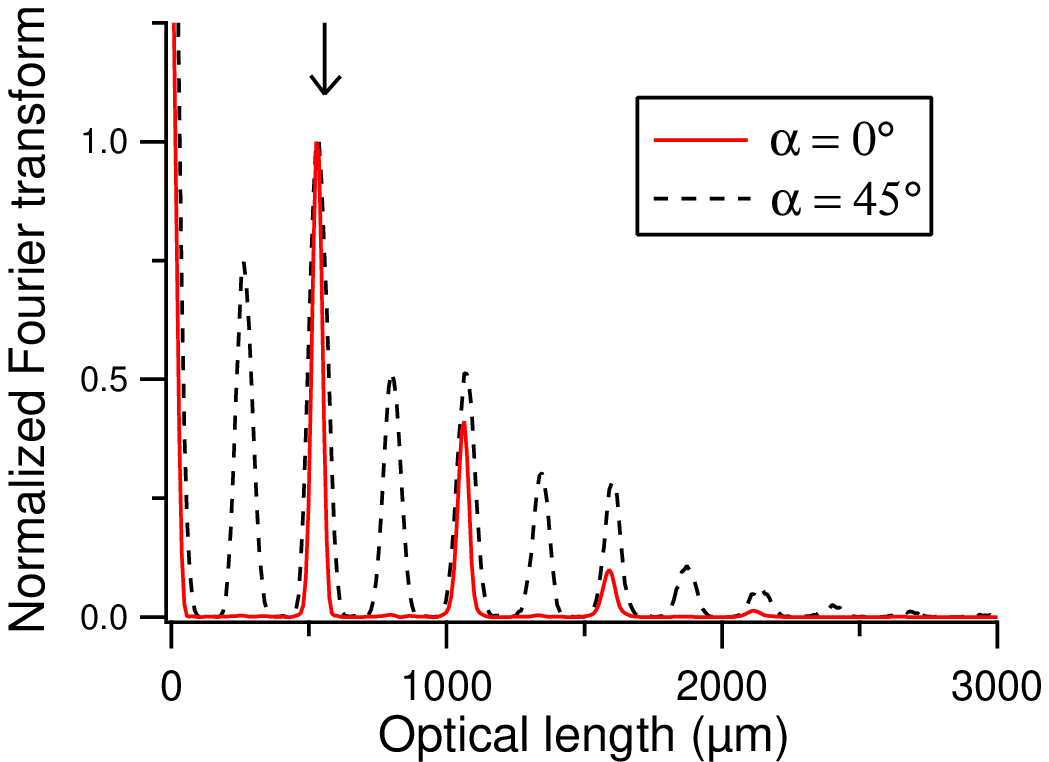}
\begin{center}(b)\end{center}
\end{minipage}\hfill
\caption{(Color online) (a) Experimental spectra from a square
micro-laser with $a=120\,\mu$m. Top: $\alpha=45^{\circ}$, bottom:
$\alpha=0^{\circ}$. (b) Fourier transform of the spectra in (a). The
arrow indicates the predicted optical length of the diamond periodic
orbit: $2\sqrt{2}a\,n$.}
    \label{fig:carre-manips}
\end{figure}

The prevalence of the diamond periodic orbit was already
experimentally demonstrated with organic micro-lasers
\cite{lebental} and micro-wave cavities \cite{bittner}. Here we
would like to stress that sometimes the experimental spectrum
reveals half the diamond periodic orbit instead of the full one due
to a selection of symmetry classes. This phenomenon is illustrated
in Fig.~\ref{fig:carre-manips}a using the pump polarization as a
control parameter. Actually the DCM molecule (the laser dye) is more
or less 'rod-like' and thus conserves (in a way) the memory of the
pump polarization which can be monitored at will without modifying
other parameters. A study of the pump polarization influence will be
published elsewhere \cite{gozhyk}.

The Fourier transforms of the spectra in
Fig.~\ref{fig:carre-manips}a are plotted in Fig.~\ref{fig:carre-manips}b.
Let us call $\alpha$ the angle between the pump polarization (which
lies in the plane of the layer) and the direction of observation.
For $\alpha=0^{\circ}$, the first harmonic of the Fourier transform
is peaked at the diamond optical length. But for
$\alpha=45^{\circ}$, only one peak out of two appears in the
spectrum, therefore the Fourier transform peaks at half the diamond
optical length. It could be noted that the second harmonic (at the
actual diamond length) is slightly higher than the first one. This
is due to the presence of a residual comb visible in the spectrum
(Fig.~\ref{fig:carre-manips}a top).

In this section, we have shown that the spectral properties of the
dielectric square (density of states, resonance losses, laser
spectra) are controlled in a first approximation by classical
features and symmetry classes. For low refractive indices (what we
studied), the diamond periodic orbit plays a prominent role.

\subsection{The dielectric rectangle}

We repeat the same steps as in the previous Section but for a
rectangular cavity so as to monitor eventual changes when breaking the square
symmetry. Let us call $\rho=L/l$ the ratio between the larger and
smaller sides. We will focus on the case $\rho=2$.

\subsubsection{Numerics}

\begin{figure}
\vspace{1cm}
\centering\includegraphics[width=.9\linewidth]{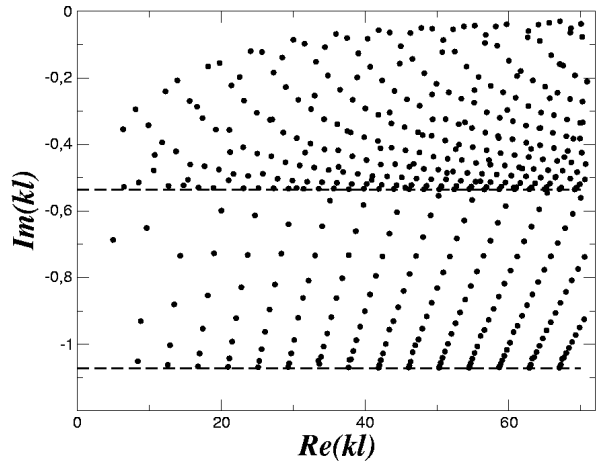}
\caption{Resonance spectrum of the dielectric rectangle with
$\rho=2$, $(++)$ symmetry class, and n=1.5. The positions of the
horizontal lines are given by (\ref{bbrect}) and (\ref{gamma2}).}
\label{specrecta2}
\end{figure}

We restrict ourselves  to the $(++)$ symmetry class
with respect to the perpendicular bisectors of the sides (see dashed area in Fig. \ref{d_l_rect}b).
Fig.~\ref{specrecta2} shows the resonance spectrum for this case.
The lower bound $\gamma_{max}$ of the imaginary parts of the
resonances is related to the lifetime of the classical orbit
bouncing off perpendicularly the longest side of the rectangle:
\begin{equation}
  \label{bbrect}
\gamma_{max}=
  \frac{1}{n}\ln\left(\frac{n-1}{n+1}\right)
\end{equation}
Another horizontal line is plotted in Fig.~\ref{specrecta2}, which
corresponds to the lifetime of the orbit bouncing off
perpendicularly the smallest side of the rectangle:
\begin{equation}
  \label{gamma2}
  \gamma_{2}=
  \frac{1}{\rho n}\ln\left(\frac{n-1}{n+1}\right)
\end{equation}

\begin{figure}
\centering\includegraphics[width=1\linewidth]{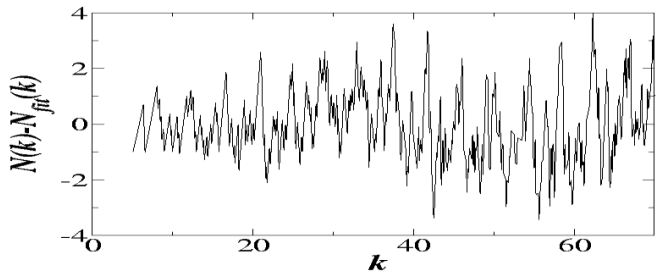}
\caption{Residue $N(k)-N_{fit}(k)$ for the dielectric rectangle with
$\rho=2$, $l=1$, $(++)$ symmetry class, and $n=1.5$.}
\label{N_k_rect}
\end{figure}

\begin{figure}
\vspace{.5cm}
  \begin{minipage}[l]{.49\linewidth}
   \centering\includegraphics[width=.9\textwidth]{fig19.eps}
    \centering (a)
  \end{minipage}
  \begin{minipage}[r]{.49\linewidth}
   \centering\includegraphics[width=.9\textwidth]{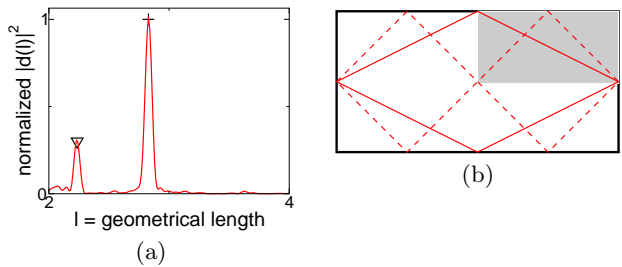}
   \vspace{1cm}
   \centering (b)
  \end{minipage}
\caption{(a) Length density for a dielectric rectangle calculated
from the spectrum in Fig. \ref{specrecta2}. The expected position of
the double diamond (resp. the stretched diamond) is indicated by a
cross (resp. a triangle). (b) Main periodic orbits contributing
  to the resonance spectrum: double diamond (dashed line) and
  stretched diamond (continuous line). The dashed area corresponds to
  the fundamental domain.}
\label{d_l_rect}
\end{figure}

As for the square cavity, prediction (\ref{eq:smooth}) is checked.
The best quadratic fit of the counting function $N(k)$ computed
numerically from data of Fig.~\ref{specrecta2} is (with $n=1.5$):
\begin{equation}
  \label{weylrect}
  N_{fit}(k)=n^2\frac{\rho}{16\pi}(kl)^2+0.2894\, kl -3.7346\ .
\end{equation}
The prediction for the linear term is:
\begin{equation}
  \label{eq:linearrect}
  \alpha^{th}= \frac{\tilde{r}(n)+n}{8\pi}(1+\rho)\Big |_{n=1.5,\rho=2}\simeq 0.3014
\end{equation}
which shows a good agreement. The difference between the numerically
computed $N(k)$ and its best quadratic fit (\ref{weylrect}) is shown
in Fig.~\ref{N_k_rect}.

The length density $d(l)$ defined by (\ref{eq:d-l}) is shown in
Fig.~\ref{d_l_rect}a, and is peaked at the lengths of the double
diamond orbit and the stretched diamond orbit (both displayed in
Fig~\ref{d_l_rect}b).

\begin{figure}
  \begin{minipage}{1\linewidth}
     \centering\includegraphics[width=.5\linewidth]{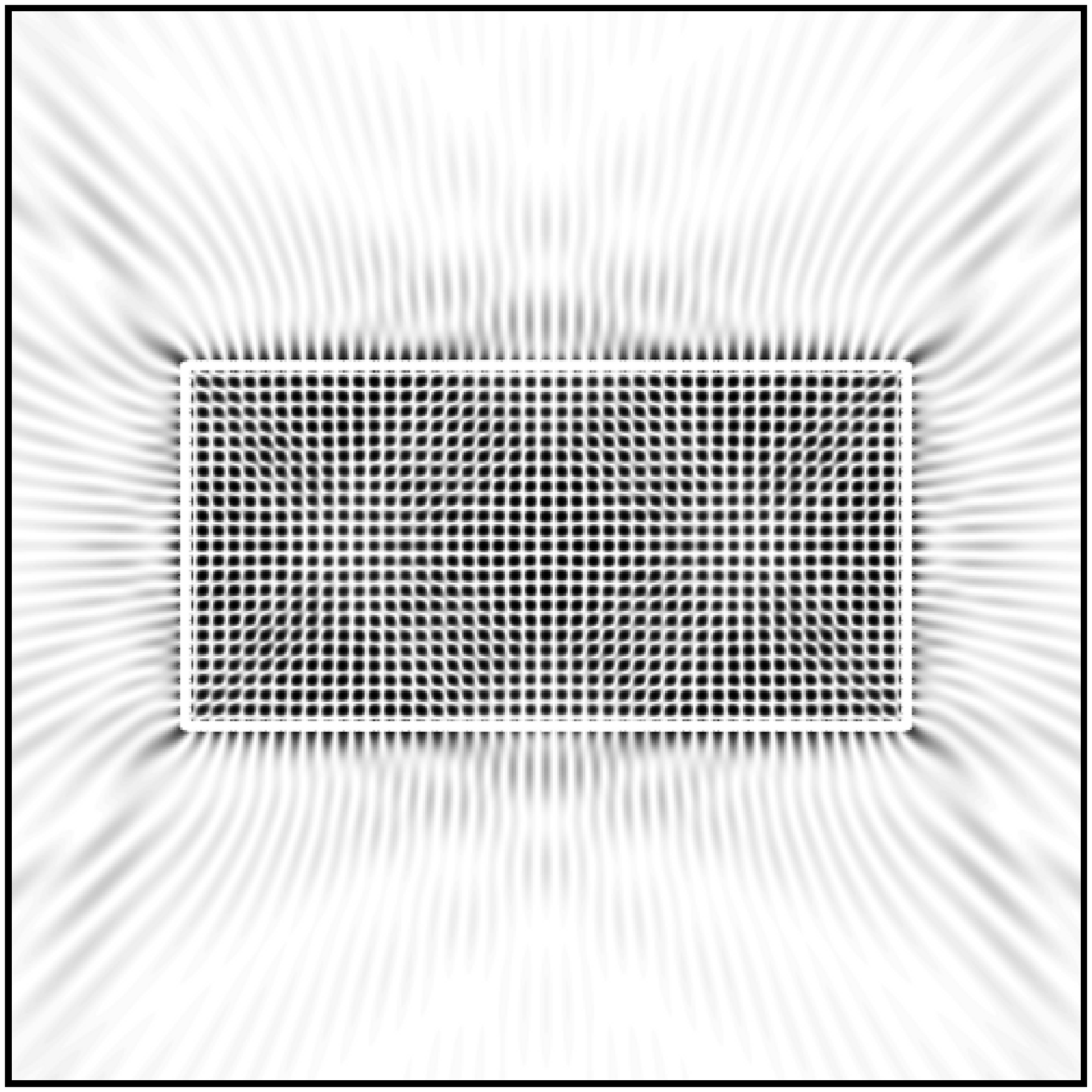}
    \centering (a)
  \end{minipage}\hfill
  \begin{minipage}{1\linewidth}
    \centering\includegraphics[width=.5\linewidth]{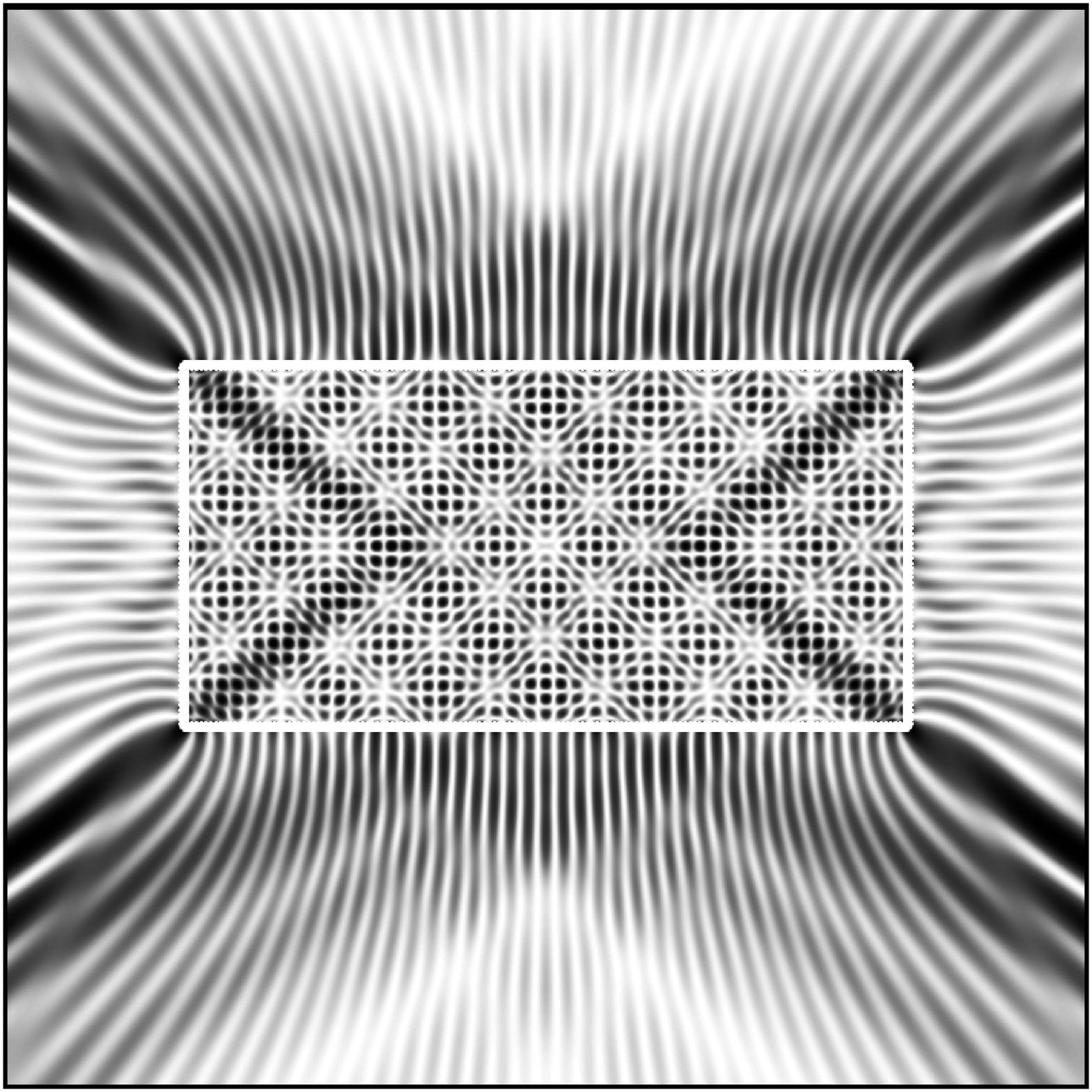}
     \centering (b)
  \end{minipage}\hfill
  \begin{minipage}{1\linewidth}
     \centering\includegraphics[width=.5\linewidth]{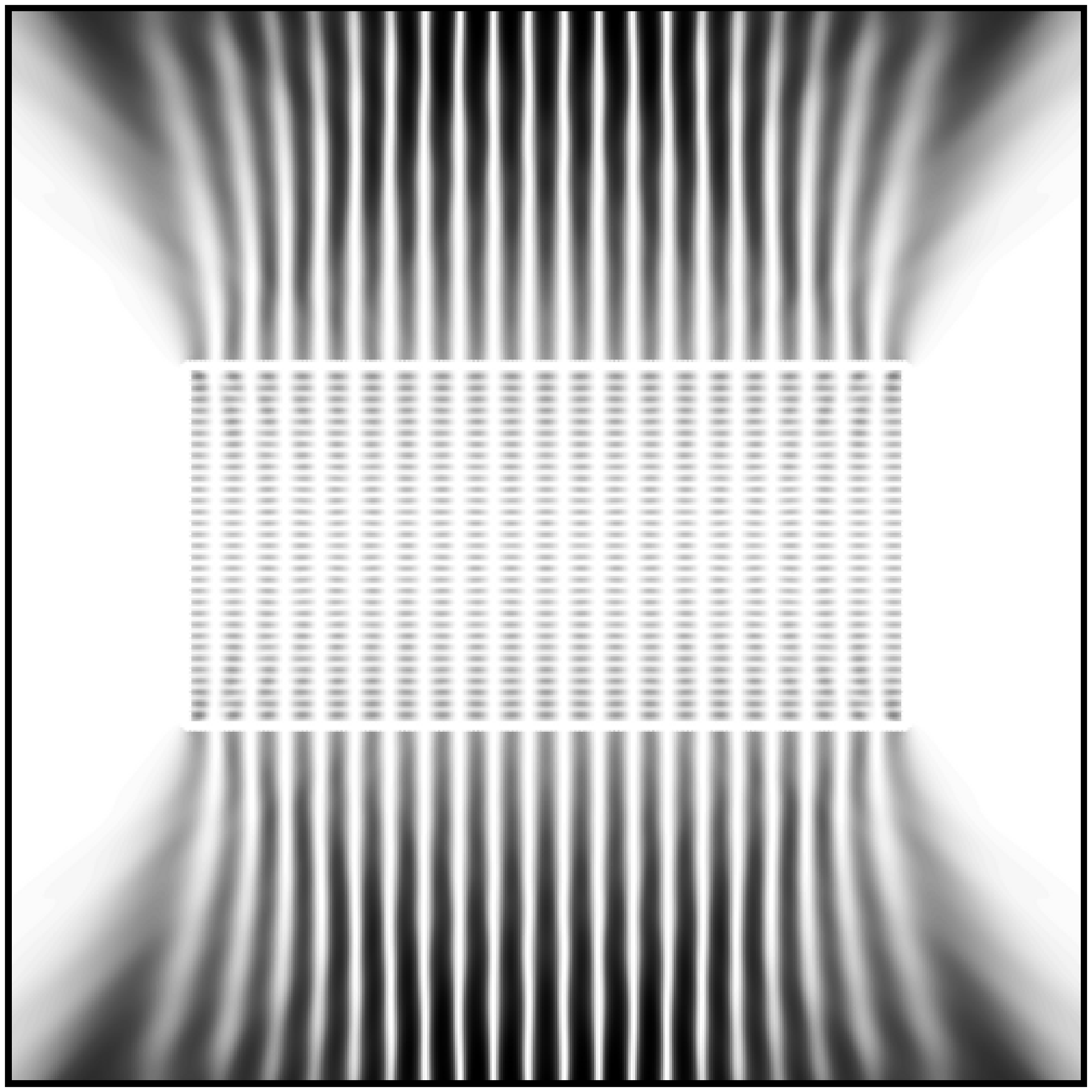}
  \centering (c)
  \end{minipage}
\caption{Quasi-stationary states for the dielectric rectangle,
  $n=1.5$, $(++)$ symmetry class. (a) $kl=70.04 -.038\,\ic$,
  (b) $kl=70.26 -.33\,\ic$. (c) $kl=70.45 -.92\,\ic$. Greyscale, black representing maximal values of $|\psi|^2$.}
  \label{psi_rect}
\end{figure}

Eventually some wave functions associated to resonances from
different parts of the spectrum are shown in Fig~\ref{psi_rect}.

\subsubsection{Experiments}

\begin{figure}
\begin{minipage}[t!]{1\linewidth}
\includegraphics[width=1\linewidth]{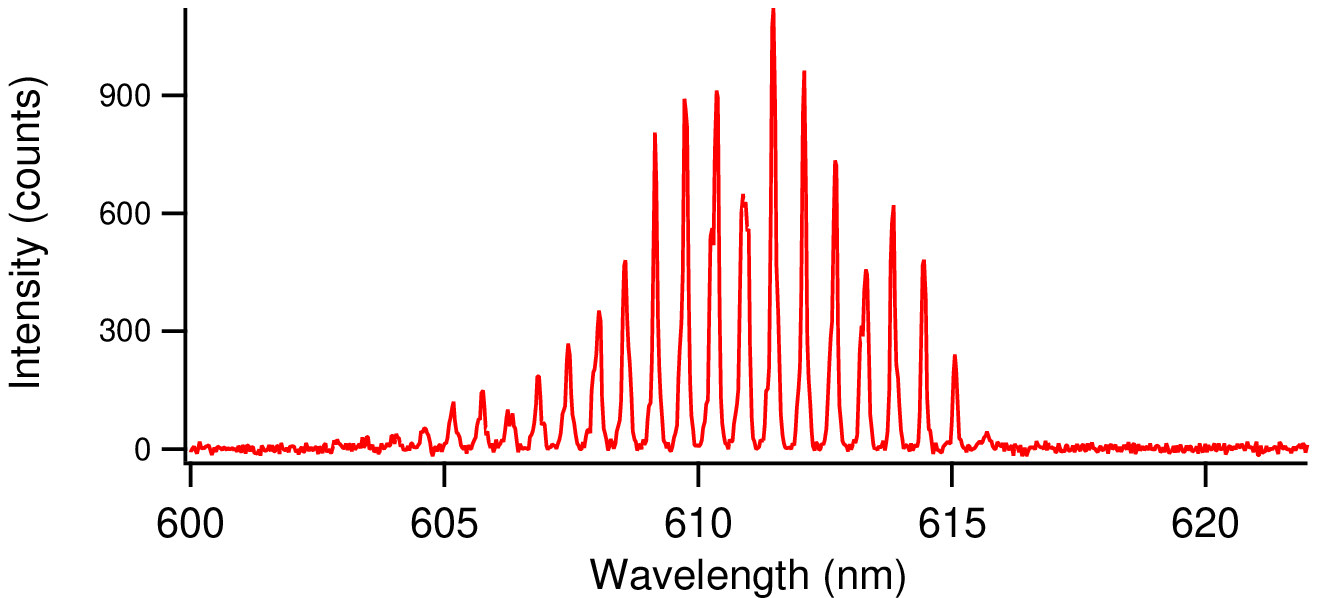}
\begin{center}(a)\end{center}
\end{minipage}\hfill
\begin{minipage}[t!]{1\linewidth}
\includegraphics[width=1\linewidth]{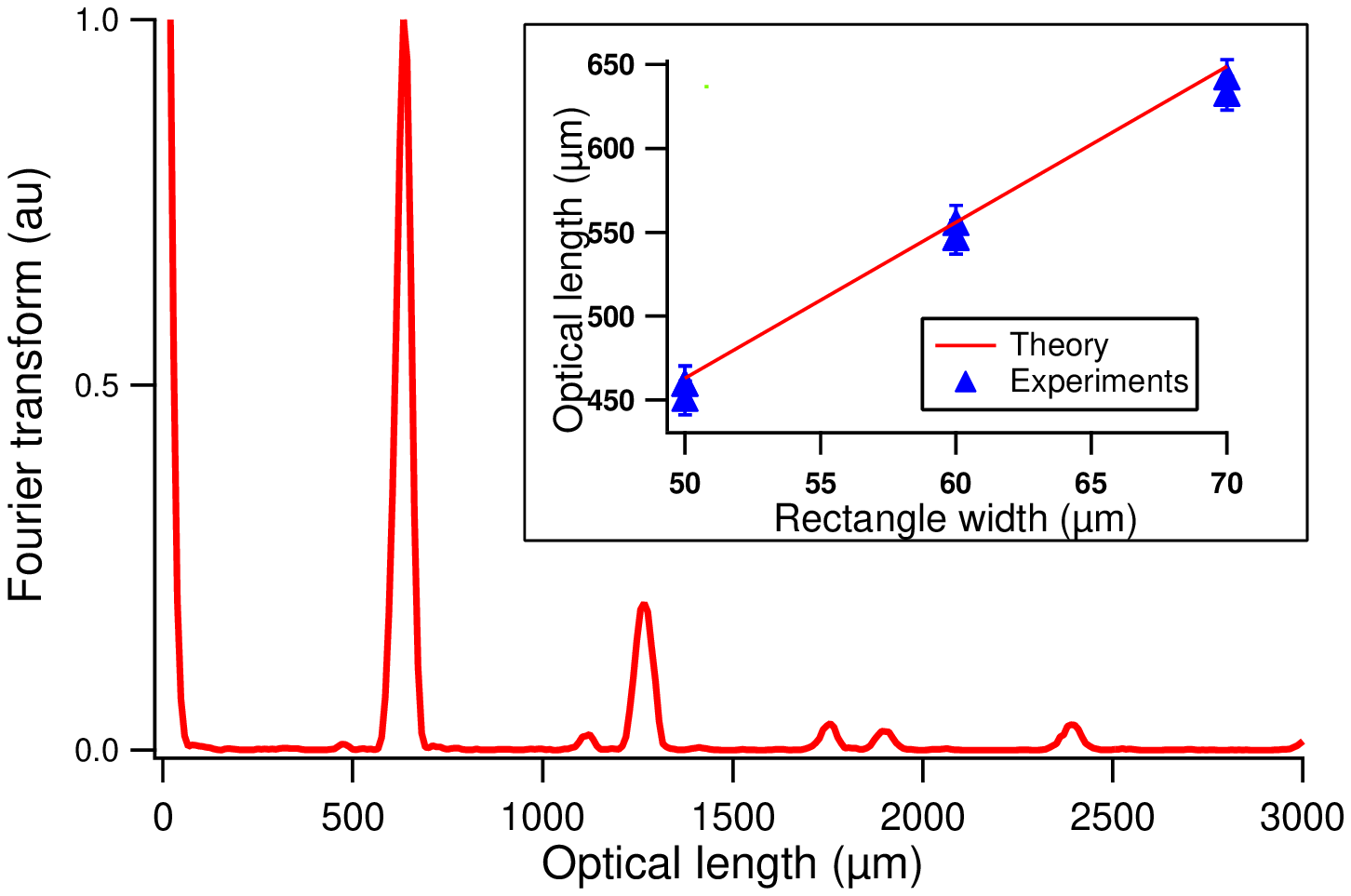}
\begin{center}(b)\end{center}
\end{minipage}\hfill
\caption{(Color online) (a) Experimental spectrum of a rectangle
microlaser with $\rho=2$ and $l=70\,\mu$m. (b) Fourier transform of
the spectrum in (a). Insert: Comparison between the optical lengths
inferred from experiments and those expected for the double-diamond
for different cavity sizes.}
    \label{fig:rectangle-spectre}
\end{figure}

Fig.~\ref{fig:rectangle-spectre}a presents a typical experimental
spectrum from a rectangular microlaser with $\rho=2$. Its Fourier
transform plotted in Fig.~\ref{fig:rectangle-spectre}b is peaked at
the length of the double diamond periodic orbit (see insert), in
agreement with numerics and predictions. This experimental
observation is very robust whatever the parameter being used:
direction of emission, pump intensity, and pump polarization. For
illustration, a comparison between the measured and expected optical
lengths is presented in insert of Fig. \ref{fig:rectangle-spectre}b
for various cavity sizes. For completeness, it should be noticed
that the Fabry-Perot along the longest axis appears if observed in
its specific direction and pumped with a favorable polarization.

\subsection{The dielectric ellipse}

The ellipse can also be considered as a 'regular shape', since the
interior billiard problem is separable \cite{bill_ell}.  Let us call
$b$ (resp. $a$) half the length of the minor (resp. major) axis.
Here we consider only the ratio $\rho\equiv a/b=2$, however the
computations for other $\rho$ values give similar results. Here we will restrict ourselves to the $(--)$
symmetry class, i.e. the function vanishes along both symmetry axis
of the ellipse.

\subsubsection{Numerics}

\begin{figure}
\centering\includegraphics[width=1\linewidth]{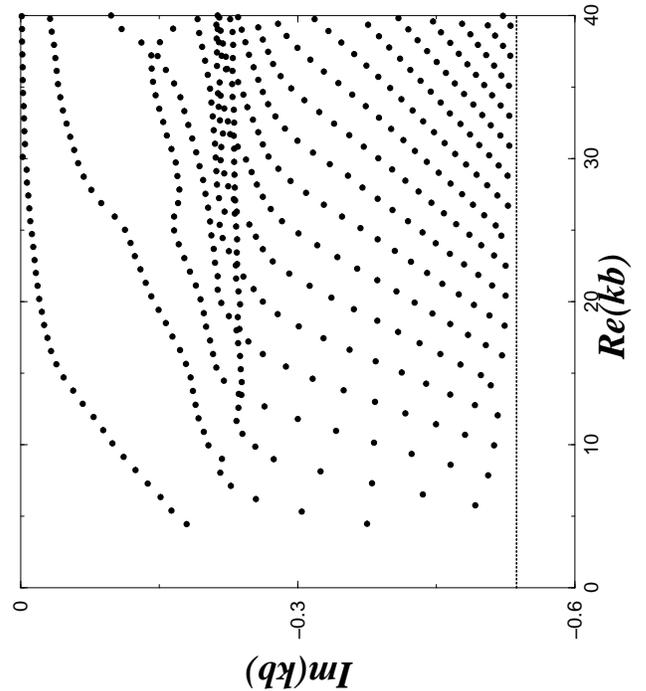}
\caption{Resonance spectrum for the dielectric ellipse with
$\rho=2$, $(--)$ symmetry class, and $n=1.5$. The position of the
horizontal dashed line is given by (\ref{bbell}).} \label{specell}
\end{figure}

Fig.~\ref{specell} shows the resonance spectrum for $\rho=2$. As for
the square cavity, it looks quite regular while the problem is not
separable. Similarly, the imaginary parts of the resonances are
bounded by the losses of the Fabry-Perot periodic orbit (along the
minor axis):
\begin{equation}
  \label{bbell}
  \quad\gamma_{max}=
  \frac{1}{2n}\ln\left(\frac{n-1}{n+1}\right)
\end{equation}

\begin{figure}
\centering\includegraphics[width=1\linewidth]{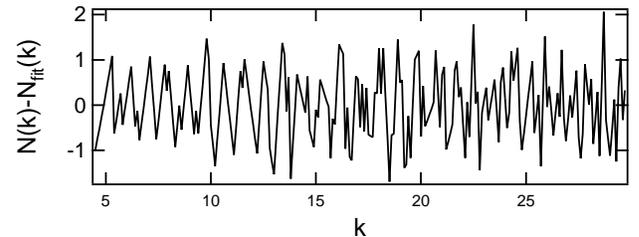}
\caption{$N(k)-N_{fit}(k)$ for the
  dielectric ellipse ($b=1$) with $\rho=2$ and $(--)$ symmetry class.}
\label{N_k_ellipse}
\end{figure}

Formula (\ref{eq:smooth}) for the counting function is checked with
the same protocol as before. The residue between numerics
and fit is plotted in Fig.~\ref{N_k_ellipse} and oscillates around
zero. Moreover the linear term of the regression
$$  \alpha^{fit}= -0.0866$$
agrees well with the prediction:
$$\alpha^{th}=\frac{\tilde{r}(n) \rho E(e)-n(1+\rho)}{4\pi}\Big|_{n=1.5,\rho=2}\simeq
  -0.0872$$
where $E(z)$ is the complete elliptic integral:
\begin{equation}
  E(z)=\int_0^{\pi/2}\sqrt{1-z^2\sin(t)^2}\ud t\ ,
\end{equation}
and $e=\sqrt{1-(b/a)^2}$ is the eccentricity of the ellipse.

\begin{figure}
  \begin{minipage}[l]{.49\linewidth}
    \includegraphics[width=.9\textwidth]{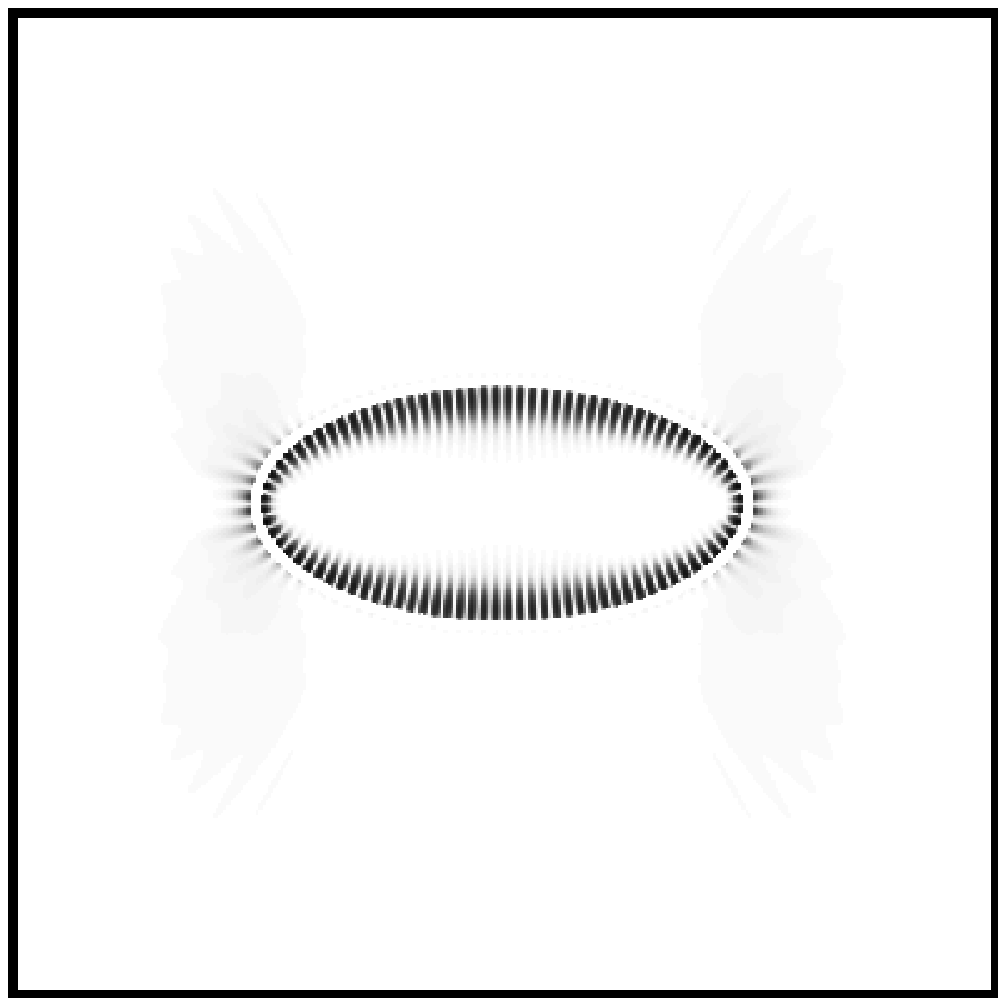}
    { \centering (a)}
  \end{minipage}
  \begin{minipage}[r]{.49\linewidth}
    \includegraphics[width=.9\textwidth]{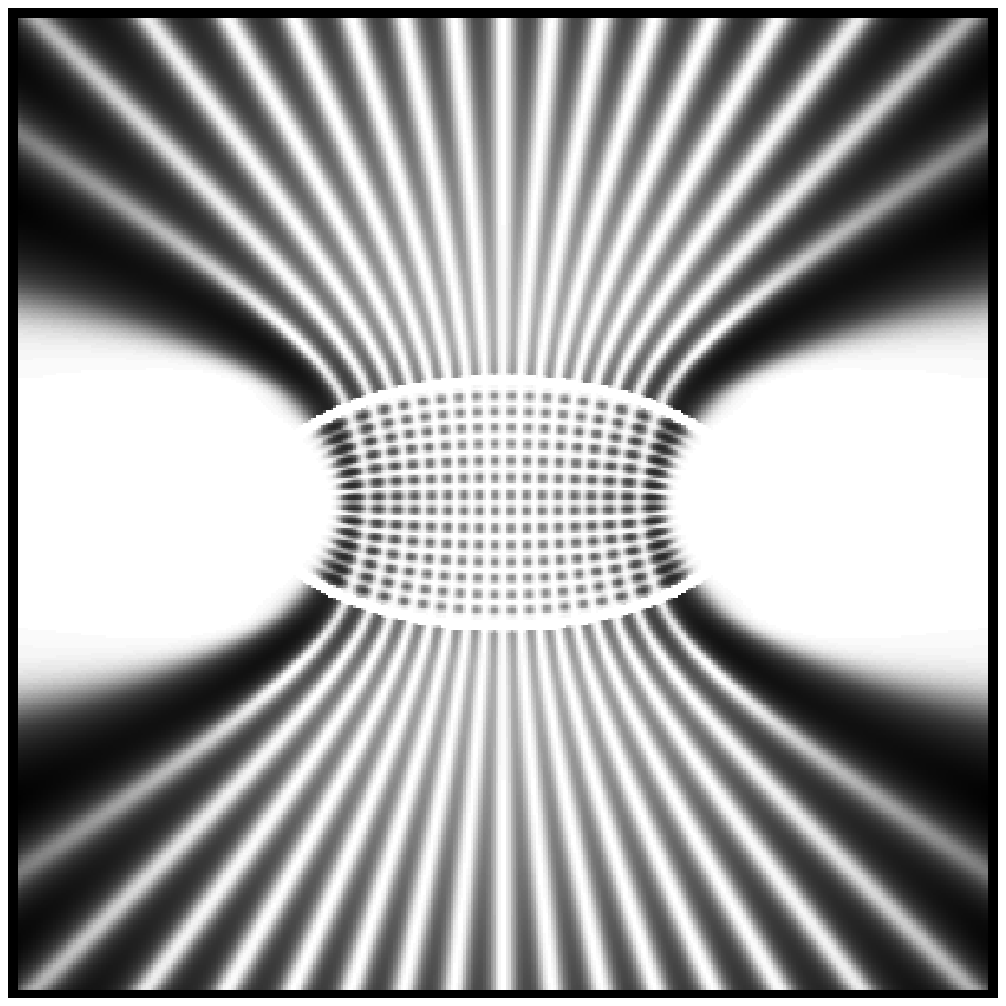}
    { \centering (b)}
  \end{minipage}
\caption{Wave functions for the ellipse with $\rho=2$, $n=1.5$ and
$(--)$ symmetry. (a) Whispering gallery mode, $kb=22.00-0.02\,\ic$.
(b) Bouncing ball mode, $kb=22.30-0.36\,\ic$. Greyscale, black
representing maximal values of $|\psi|^2$.} \label{fig:psi-ell}
\end{figure}

The oscillatory part will be postponed to a future publication. We
already note that the wave functions display in general two kinds of
behavior for the wave inside the cavity: either ``whispering gallery
modes'' or ``bouncing ball modes'' - like as is (rigorously) the
case for the elliptic billiard. Fig. \ref{fig:psi-ell} presents
examples of such wave functions, which correspond to resonances from
different parts of the spectrum.

\subsubsection{Experiments}

\begin{figure}
\includegraphics[width=1\linewidth]{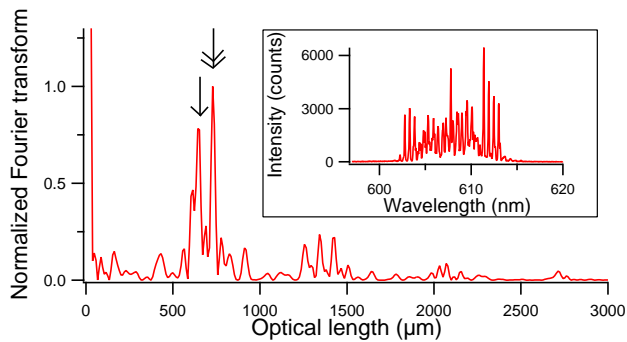}
\caption{Fourier transform of the spectrum in insert. The arrows
indicate the predicted optical length of periodic orbits: single
arrow for the Fabry-Perot along the major axis $4a\,n$ and double
arrow for the rectangle $4bn\sqrt{1+\rho^2}$. Insert:
Experimental spectrum from an elliptical micro-laser with
$\rho=a/b=2$ and $b=50\,\mu$m registered in the direction parallel
to the major axis.}
\label{fig:ellipse-manips}
\end{figure}

Experiments provide similar insights onto the dominant resonance
features. The insert in Fig. \ref{fig:ellipse-manips} shows a
typical spectrum from an elliptical micro-laser with $\rho=2$, while
its Fourier transform is plotted in the main window. Its first
harmonics presents two main peaks with positions corresponding quite
well to the optical lengths of two periodic orbits: the rectangle
and the Fabry-Perot along the major axis. The deviation is less than
3 \%, which is the experimental inaccuracy. It should be noted that for $\rho=2$ the length of
the Fabry-Perot along the major axis is equivalent to the second repetition
of the Fabry-Perot along the minor axis, also called bouncing ball.

\section{Pseudo-integrable system: the dielectric pentagon}\label{third}

Similar studies were performed for the dielectric pentagon and
hexagon, which are particularly interesting systems since they
contain diffracting angles: $m\pi/n$ with $m,n$ co-prime and $m>1$.
Following Richens and Berry \cite{pseudoint} these systems are
called pseudo-integrable, since their classical flow is confined
to a surface as for integrable systems but  because its genus is  bigger than 1
they cannot be classically integrable.

Here we only consider the dielectric pentagon, though every
conclusion also applies to the hexagon \emph{mutatis mutandis}.
Below we present the results for the $(--)$ symmetry class, which
means that the associated wave functions vanish along each symmetry
axis of the polygon. $R$ stands for the radius of the outer circle
of the pentagon and $a=2R\sin(\pi/5)$ for its side length.

\subsection{Numerics}

\begin{figure}
  \begin{minipage}[l]{.99\linewidth}
      \includegraphics[width=.99\textwidth]{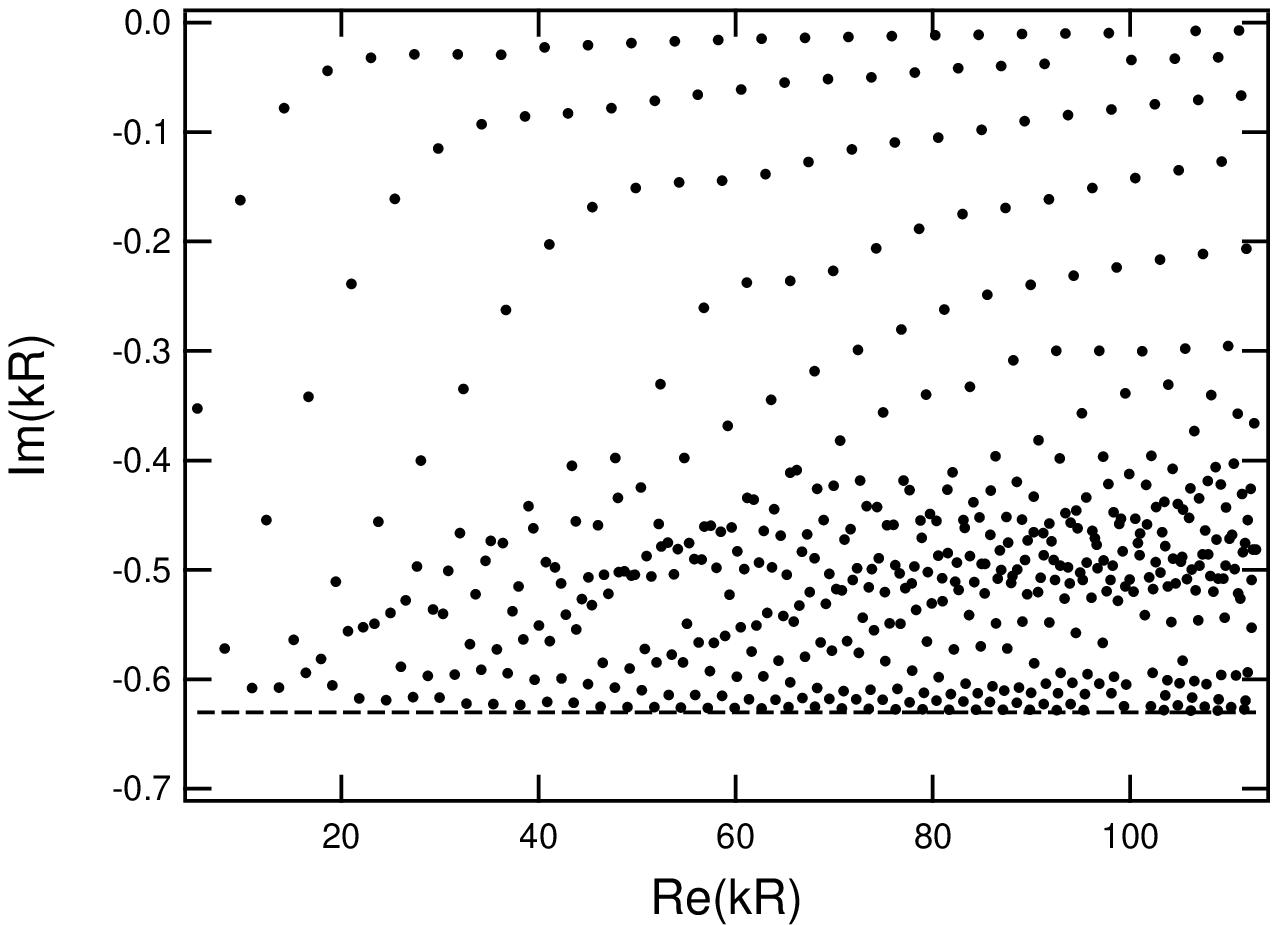}
    {\centering (a)}
  \end{minipage}
   \begin{minipage}[r]{1\linewidth}
\includegraphics[width=1\textwidth]{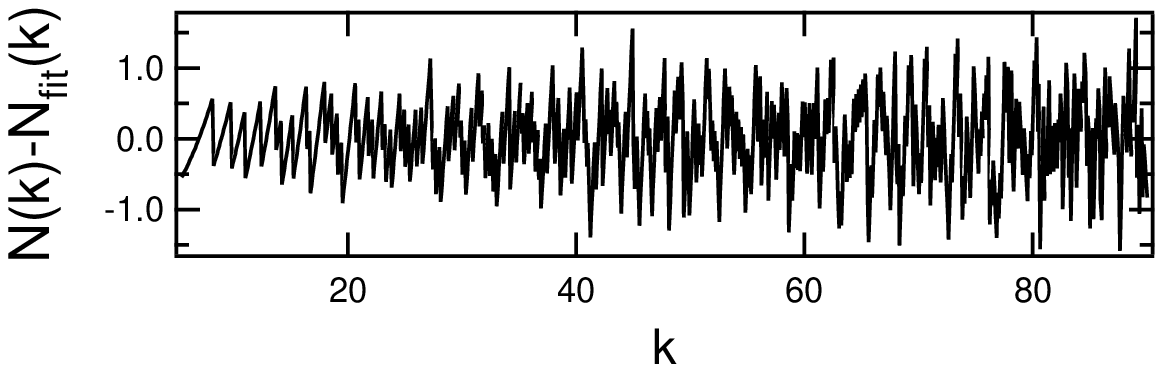}
    {\centering (b)}
  \end{minipage}
   \begin{minipage}[r]{.4\linewidth}
\includegraphics[width=1\textwidth]{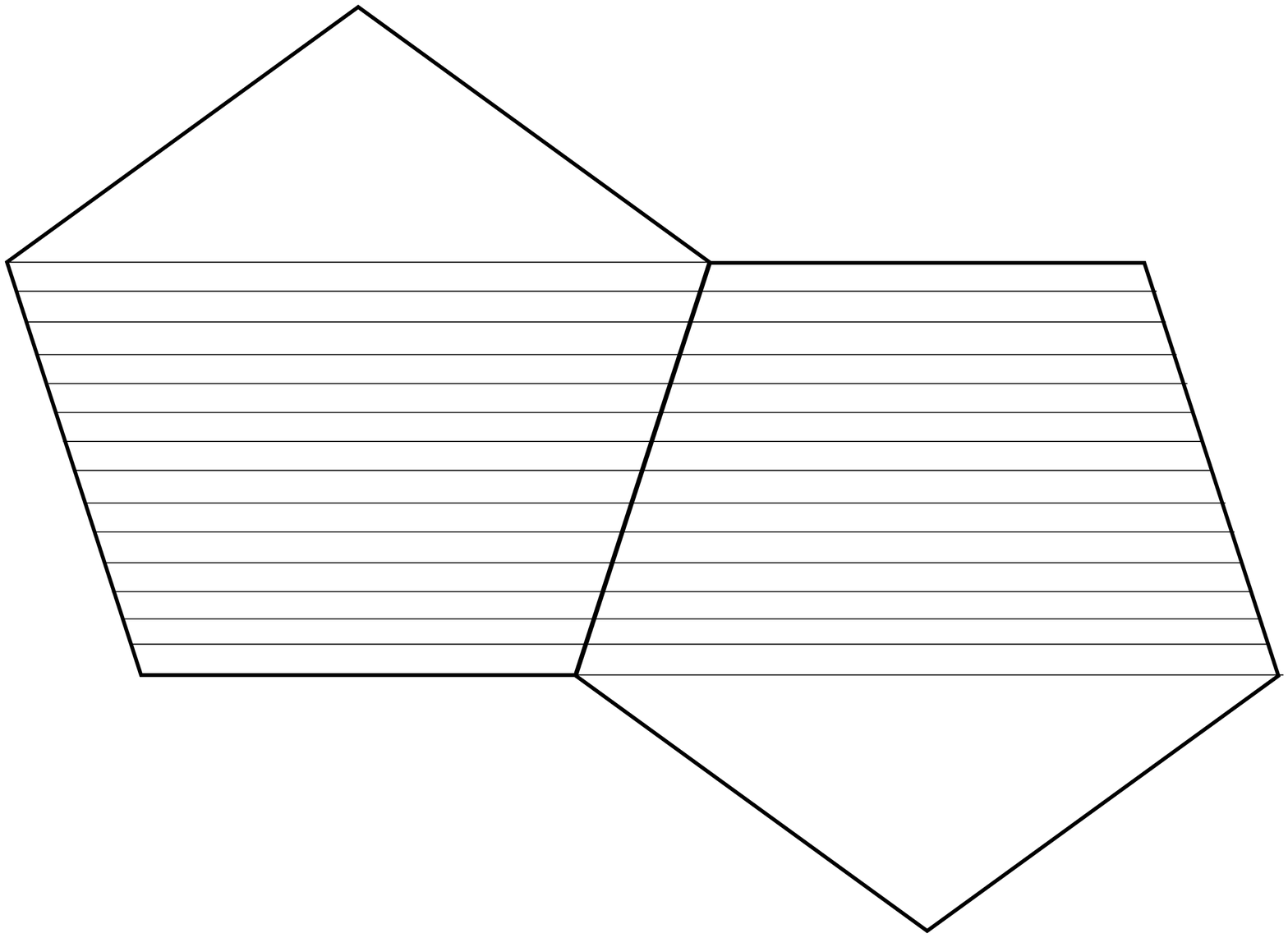}
    { \centering (c)}
  \end{minipage}
   \begin{minipage}[r]{.4\linewidth}
\includegraphics[width=0.7\textwidth]{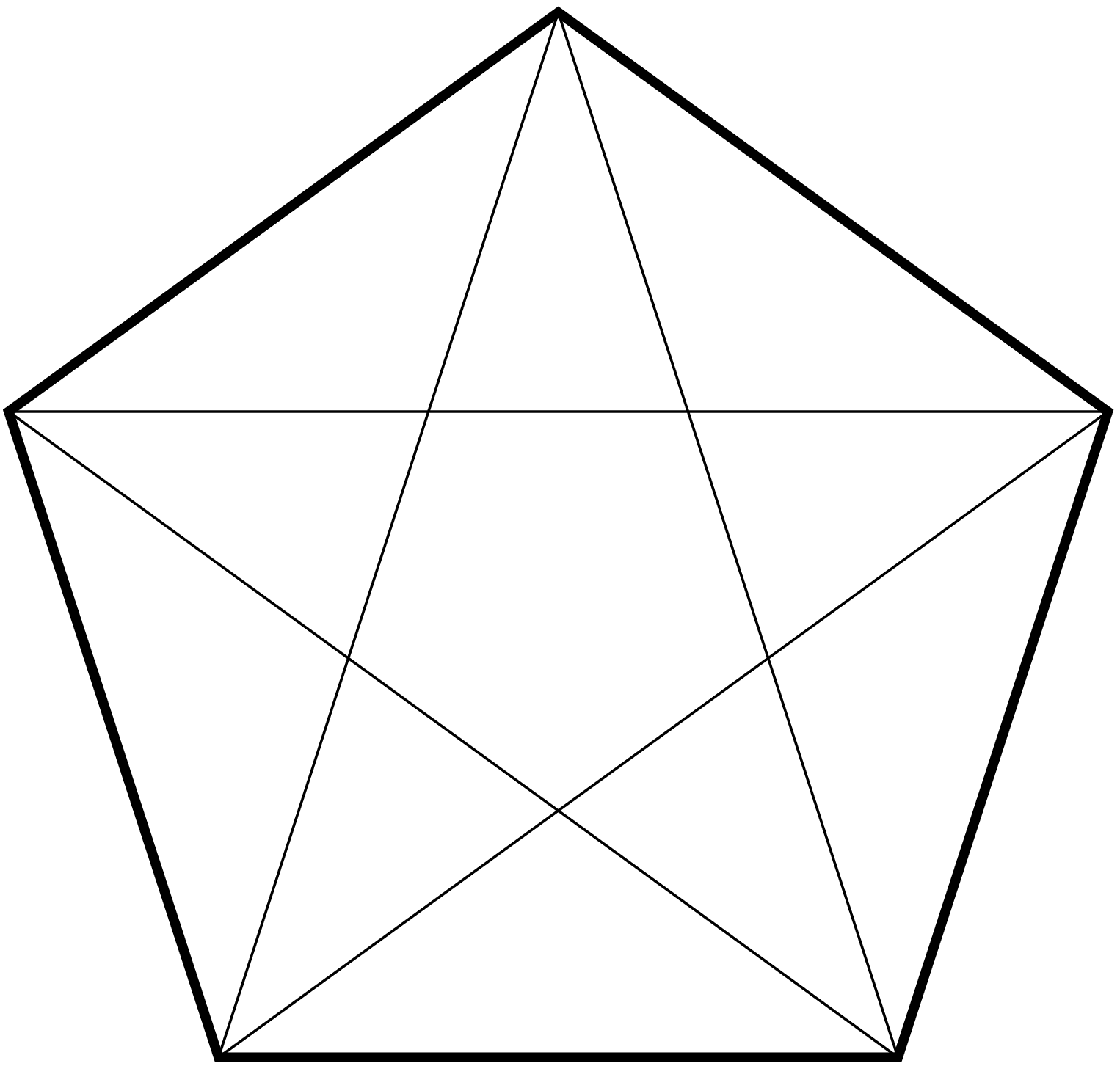}
\begin{center}(d)\end{center}
  \end{minipage}
\caption{(a) Resonance spectrum for the dielectric pentagon with
$(--)$ symmetry and $n=1.5$. The position of the horizontal dashed
line is given by (\ref{gpenta}). (b) $N(k)-N_{fit}(k)$ calculated
from the numerical spectrum in (a) with $R=1$. (c) A part of the
periodic orbit with the shortest lifetime. (d) A representation of
the periodic orbit in (c).} \label{fig:specpenta}
\end{figure}
Fig.~\ref{fig:specpenta}a shows the resonance spectrum of a
dielectric pentagon for $(--)$ symmetry and $n=1.5$. Again the
imaginary part of the resonances is bounded and this lower bound
$\gamma_{max}$ can be estimated from the refractive losses of the
periodic orbit drawn in Fig.~\ref{fig:specpenta}c and d which
presents the highest losses (shortest lifetime):
\begin{equation}
  \label{gpenta}
   \gamma_{max}=\frac{2}{nl}\ln R_{TM}\left(\frac{\pi}{10}\right)\Big |_{n=1.5}\simeq
   -0.630
\end{equation}
where $l=2R\left(\sin(2\pi/5)+\sin(\pi/5)\right)$.

\begin{figure}
  \begin{minipage}[l]{1\linewidth}
\includegraphics[width=.4\linewidth,angle=-90]{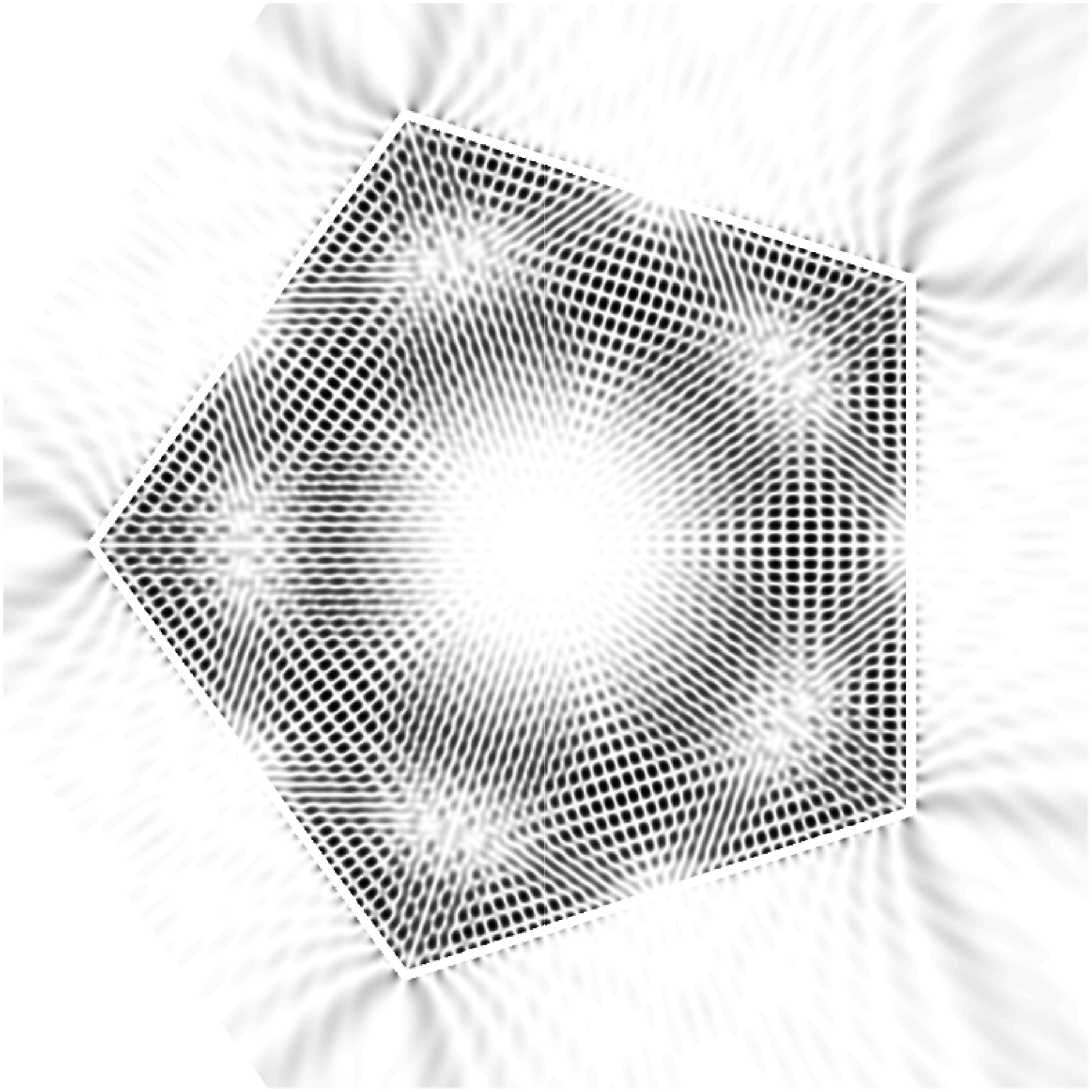}
    \centering (a)
  \end{minipage}
  \begin{minipage}[r]{1\linewidth}
    \includegraphics[width=.4\linewidth,angle=-90]{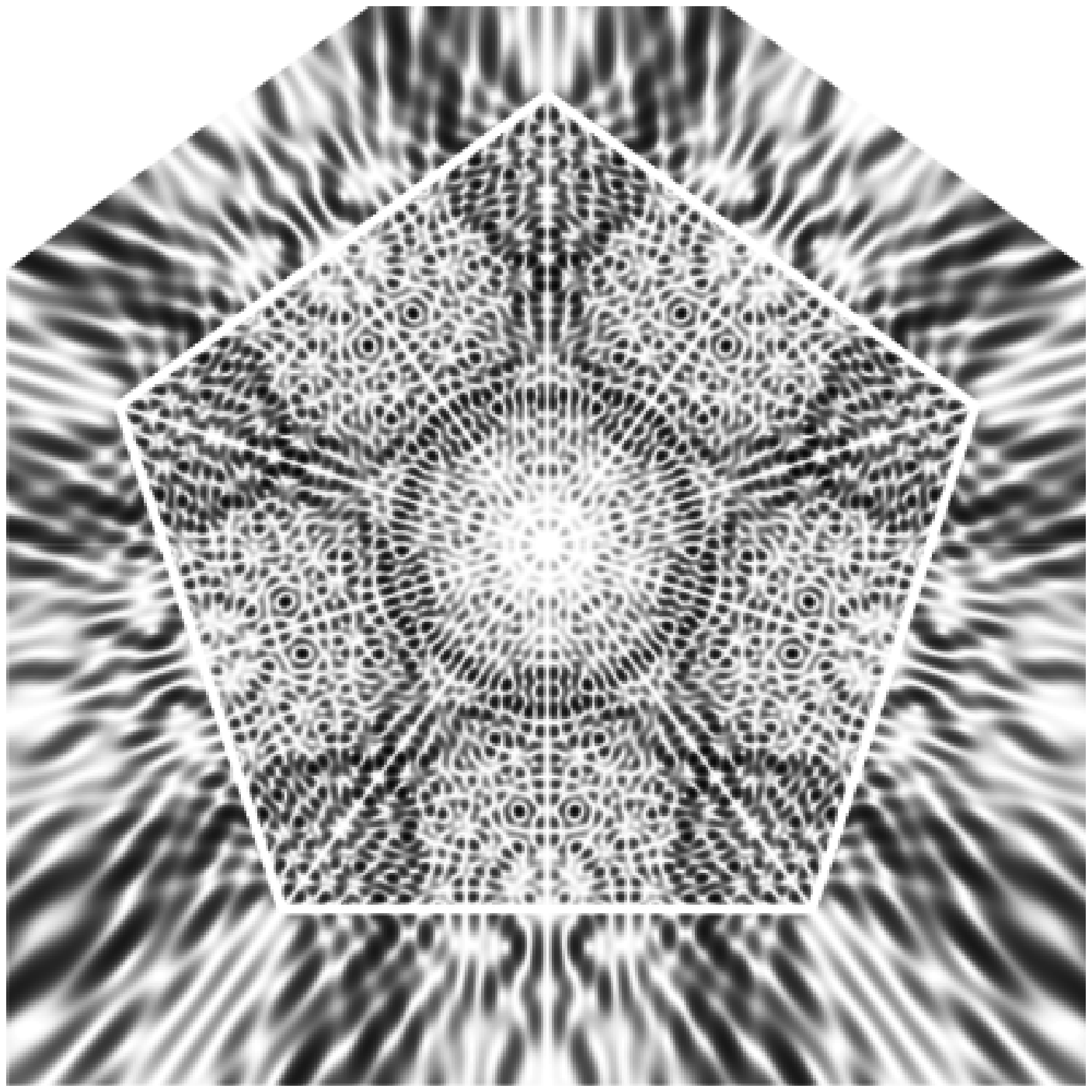}
    \centering (b)
  \end{minipage}
\begin{minipage}[r]{1\linewidth}
    \includegraphics[width=.4\linewidth,angle=-90]{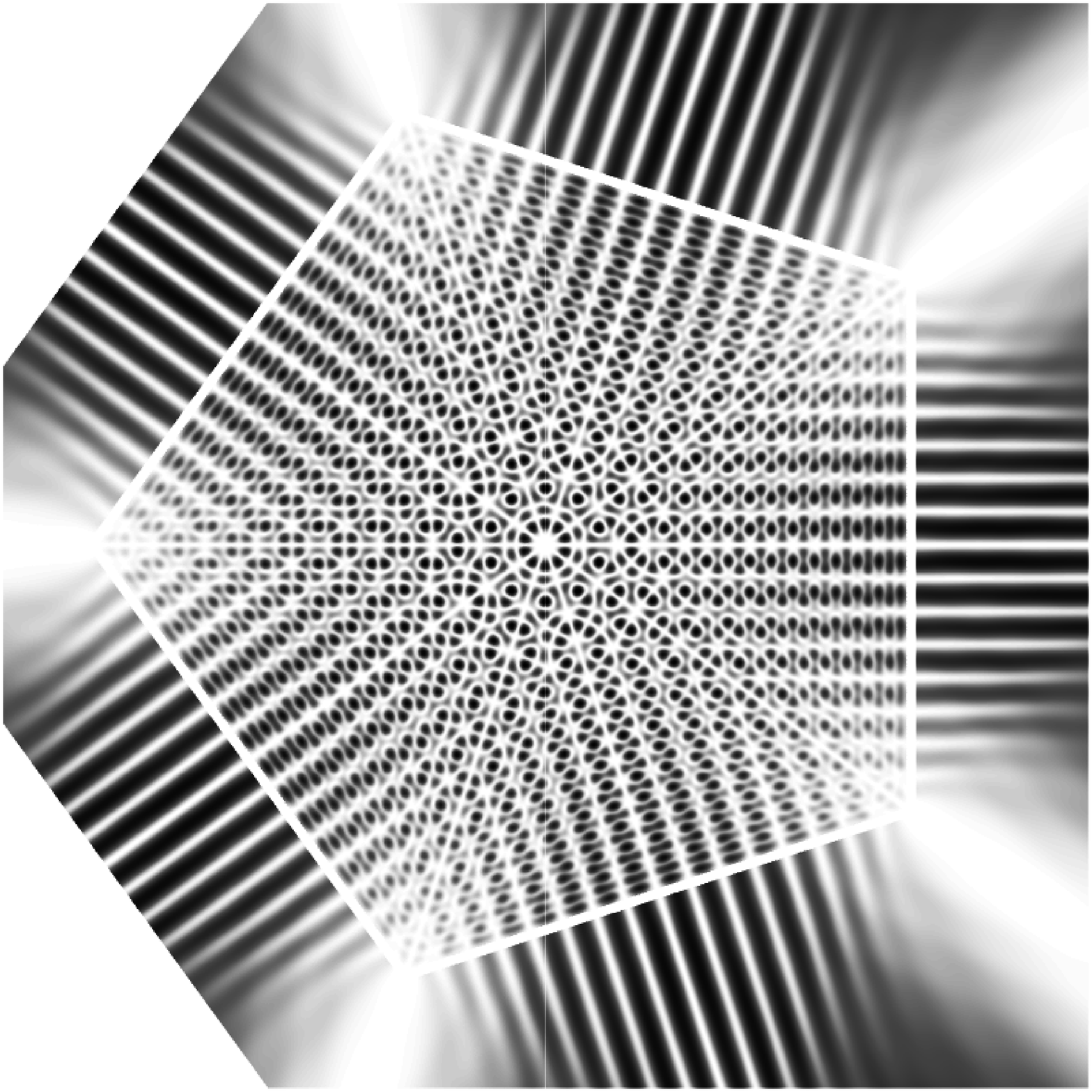}
    \centering (c)
  \end{minipage}
\caption{Wave functions of the dielectric pentagon with $(--)$
  symmetry and $n=1.5$. (a) $kR=91.35-0.038\,\ic$, (b)
  $kR=89.06-0.453\,\ic$, and (c) $kR=92.56-0.628 \,\ic$. Greyscale, black representing maximal values of $|\psi|^2$.}
\label{fig:psipenta}
\end{figure}

Several wave functions are
displayed in Fig.~\ref{fig:psipenta}. It is important to stress the existence of different types of
resonances. The ones with low losses are related to 'whispering
gallery-like' modes, see Fig.~\ref{fig:psipenta}a. In
\cite{lebental}, this observation was used to build a superscar
approximation of these resonances. Notice the more complex
pattern in (b) for this wave function corresponding to $kR$ with
rather large imaginary part, and the scarring by the orbit in Fig.
\ref{fig:specpenta}d for the wave function in Fig.
\ref{fig:psipenta}c.

The Weyl law (\ref{eq:smooth}) is checked as above by fitting the
counting function $N(k)$ which gives:
$$ N_{fit}(k)=\frac{n^2\sin(2\pi/5) }{16\pi} (kR)^2
  -0.1681 \,kR+0.1077 \ ,$$
in good agreement with the prediction for the linear term:
$$\alpha^{th}=\frac{\tilde{r}(n)\sin(\pi/5)-n\,[1+\cos(\pi/5)]}{4\pi} \Big|_{n=1.5}\simeq
-0.1680$$ Moreover the residue $N(k)-N_{fit}(k)$ oscillates
around zero as expected (see Fig. \ref{fig:specpenta}b).

\begin{figure}
  \begin{minipage}[l]{0.4\linewidth}
    \includegraphics[width=0.7\linewidth]{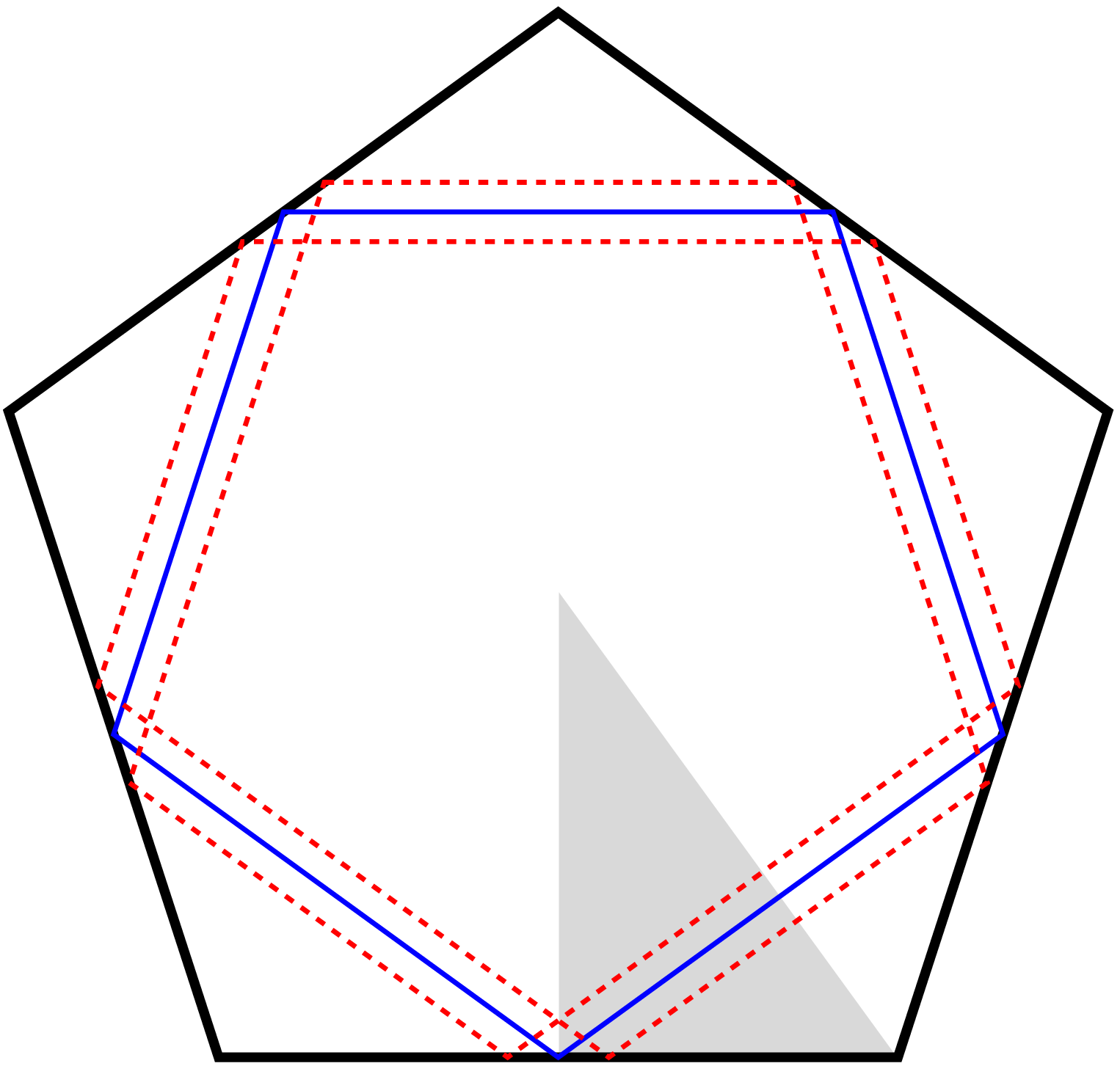}
\begin{center}(a)\end{center}
  \end{minipage}
   \begin{minipage}[r]{.4\linewidth}
\includegraphics[width=0.7\textwidth]{fig34.eps}
\begin{center}(b)\end{center}
  \end{minipage}
  \begin{minipage}[r]{\linewidth}
    \includegraphics[width=1\textwidth]{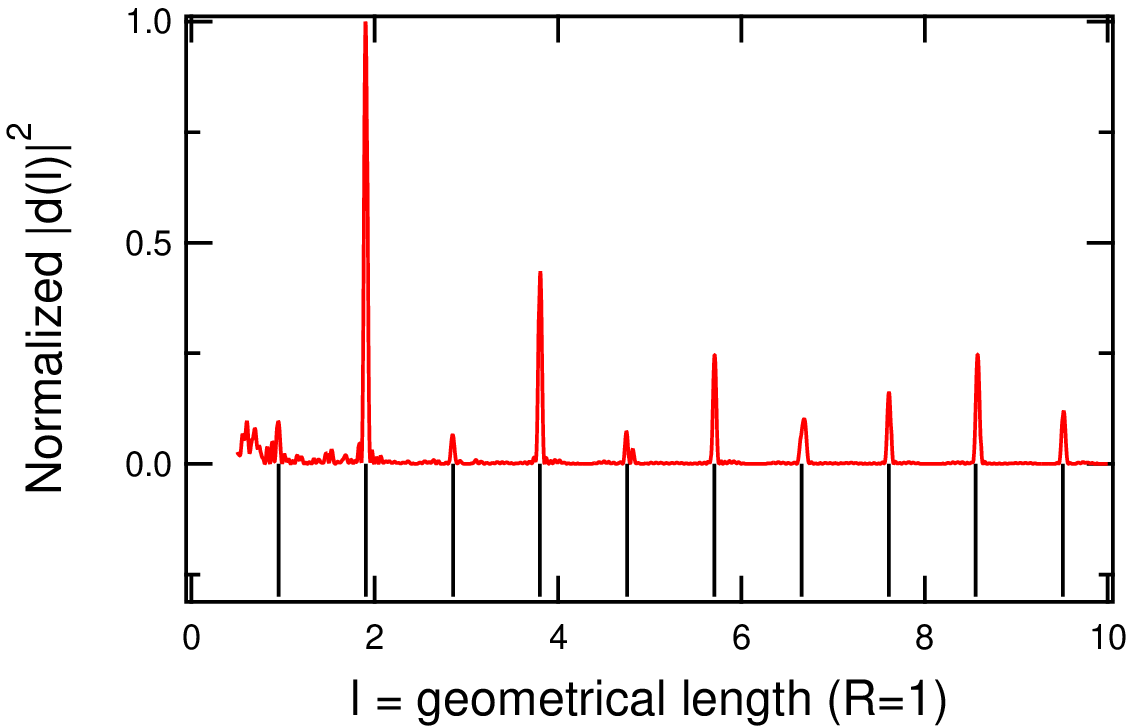}
\begin{center}(c)\end{center}
  \end{minipage}
\caption{(Color online) (a) 'single pentagon' periodic orbit in
solid line and 'double pentagon' periodic orbit in dotted line. The
dashed area indicates the fundamental domain used for simulations.
(b) A representation of the double pentagon periodic orbit. (c)
Length density for the dielectric pentagon calculated from the
numerical data plotted in Fig. \ref{fig:specpenta}a. The vertical
lines in the bottom part of the graph indicate the repetitions of the
single pentagon orbit.} \label{fig:orbpenta}
\end{figure}

The length density plotted in Fig.~\ref{fig:orbpenta}c evidences
that a few periodic orbits mostly contributes to the oscillatory
part of the trace formula. First the density of orbit length is
peaked at a length corresponding to the 'double pentagon' periodic
orbit, which is depicted in Fig.~\ref{fig:orbpenta}a and b. This
orbit is confined by total internal reflection for our value of the
refractive index and lives in family contrary to the isolated 'single
pentagon' orbit.  The length of the
single pentagon periodic orbit, once folded in the fundamental
domain shown in Fig.~\ref{fig:orbpenta}a, is $l=R\sin(2\pi/5)$. The
length of the double pentagon periodic orbit is twice longer. The
vertical lines in Fig.~\ref{fig:orbpenta}b indicate the theoretical
lengths of the $m^{th}$ repetition of the single pentagon periodic
orbit: $l_m=m\,l$. As expected, the length density is
mostly peaked at the $l_m$ positions with even $m$.

\begin{figure}
  \begin{minipage}[l]{.49\linewidth}
    \includegraphics[width=.9\textwidth]{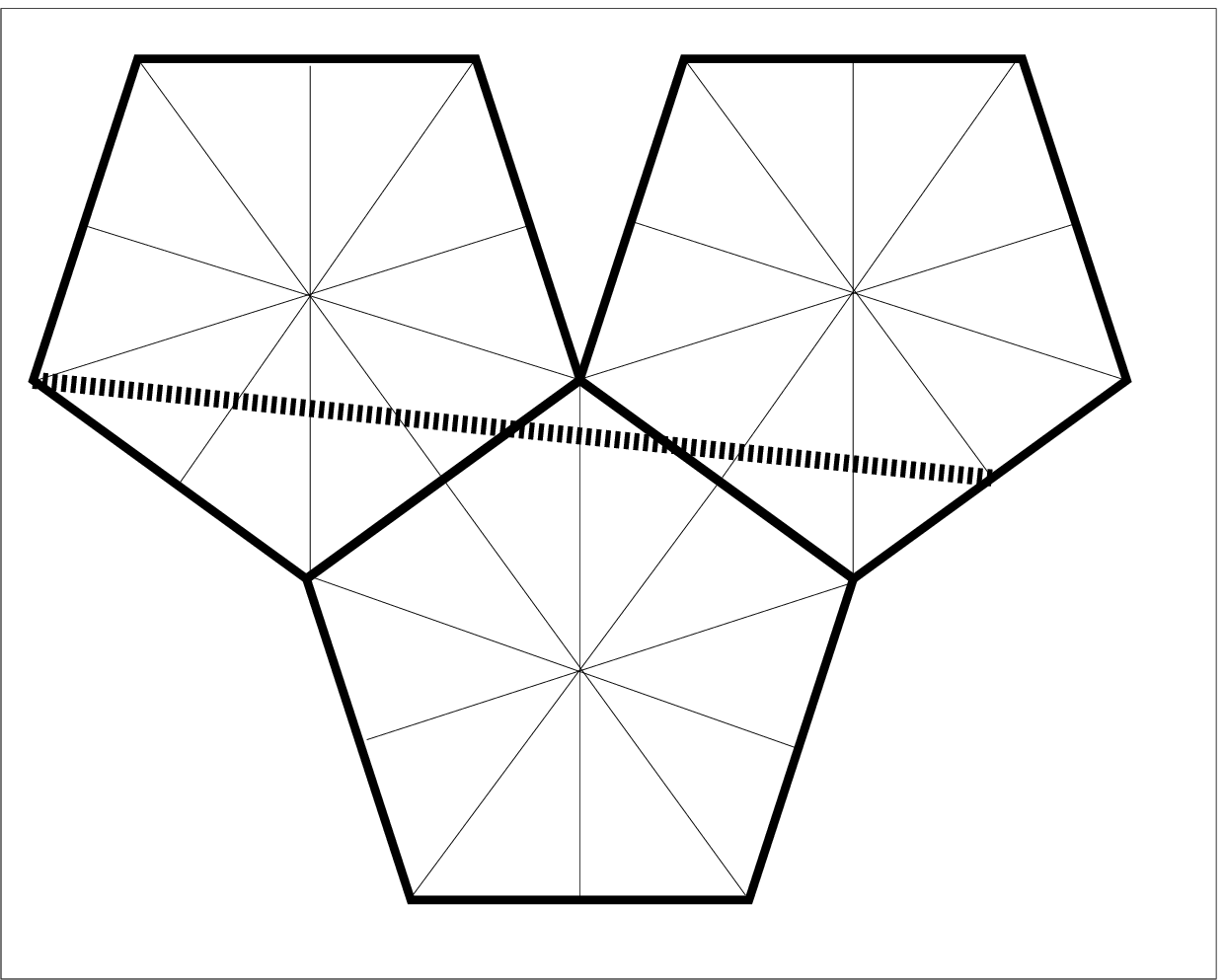}
    { \centering a)}
  \end{minipage}
  \begin{minipage}[r]{.49\linewidth}
    \includegraphics[width=.9\textwidth]{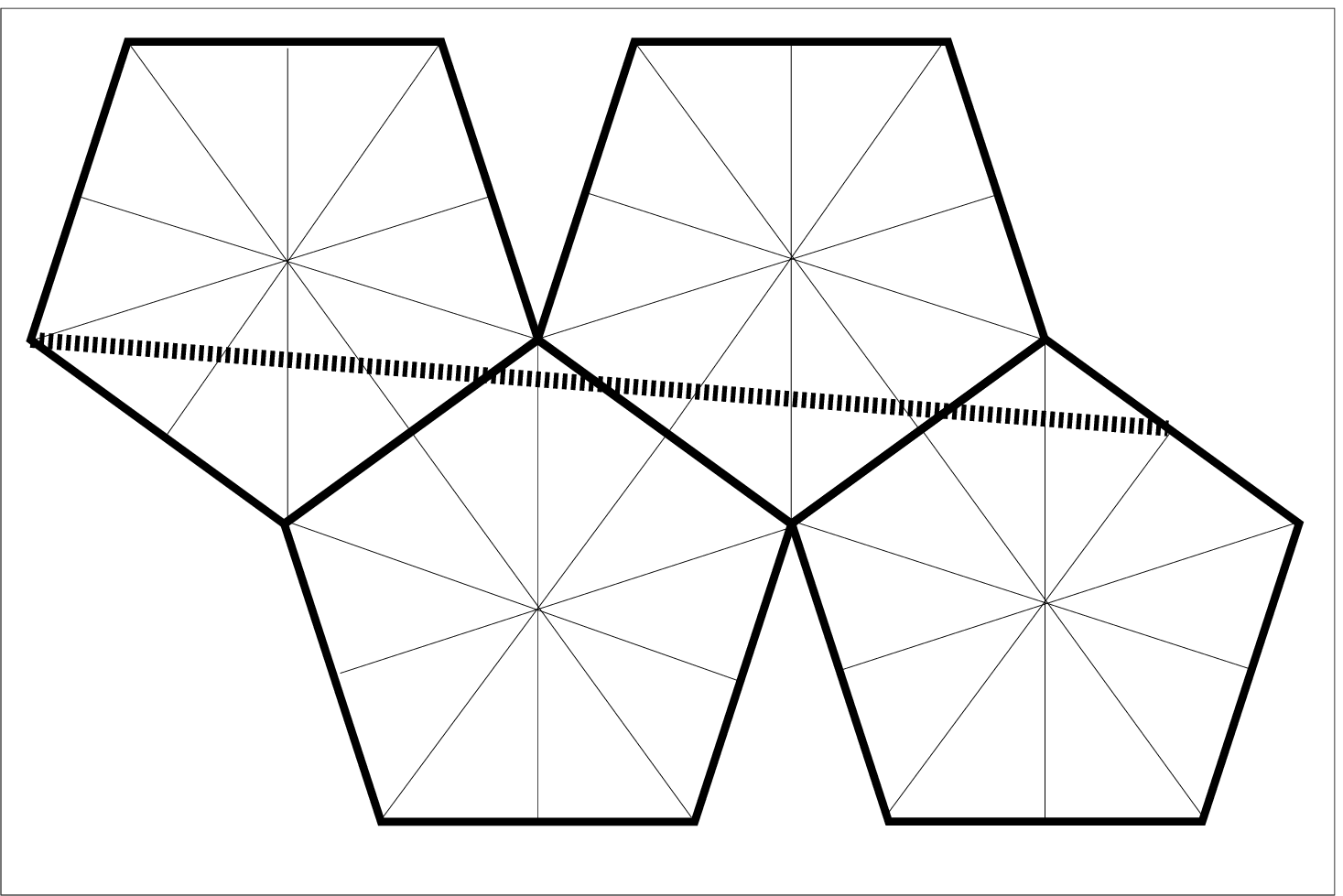}
\vspace{.5cm}
    { \centering b)}
  \end{minipage}
\caption{Diffracting orbits in the pentagonal cavity. \\
(a)  $l_a=R\sqrt{130+22\sqrt{5}}/2$, $7l/l_a\approx .995$.\\
(b) $l_b=R\sqrt{210+38\sqrt{5}}/2$, $9l/l_b\approx .997$.}
\label{orbdiffpenta}
\end{figure}

Second, it is worth noting that the amplitude of the peaks does not
clearly decay as for the square in Fig.~\ref{fig:d-l-carre}. For
$m=7$ for instance, the peak amplitude is unexpectedly high for a
repetition of a given orbit. This happens when the length $l_m$ is
close to the length of a diffracting orbit. Here, the repetitions
$m=7$ and $m=9$ are in fact of similar lengths than the orbits
illustrated in Fig.~\ref{orbdiffpenta}. The treatment of
such diffracting correction (see a similar discussion in
\cite{olivier} for billiards) is beyond the scope of this paper as
it requires the local exact solution of the diffracted field by a
dielectric wedge.

\subsection{Experiments}

\begin{figure}
\begin{minipage}[t!]{.8\linewidth}
\includegraphics[width=1\linewidth]{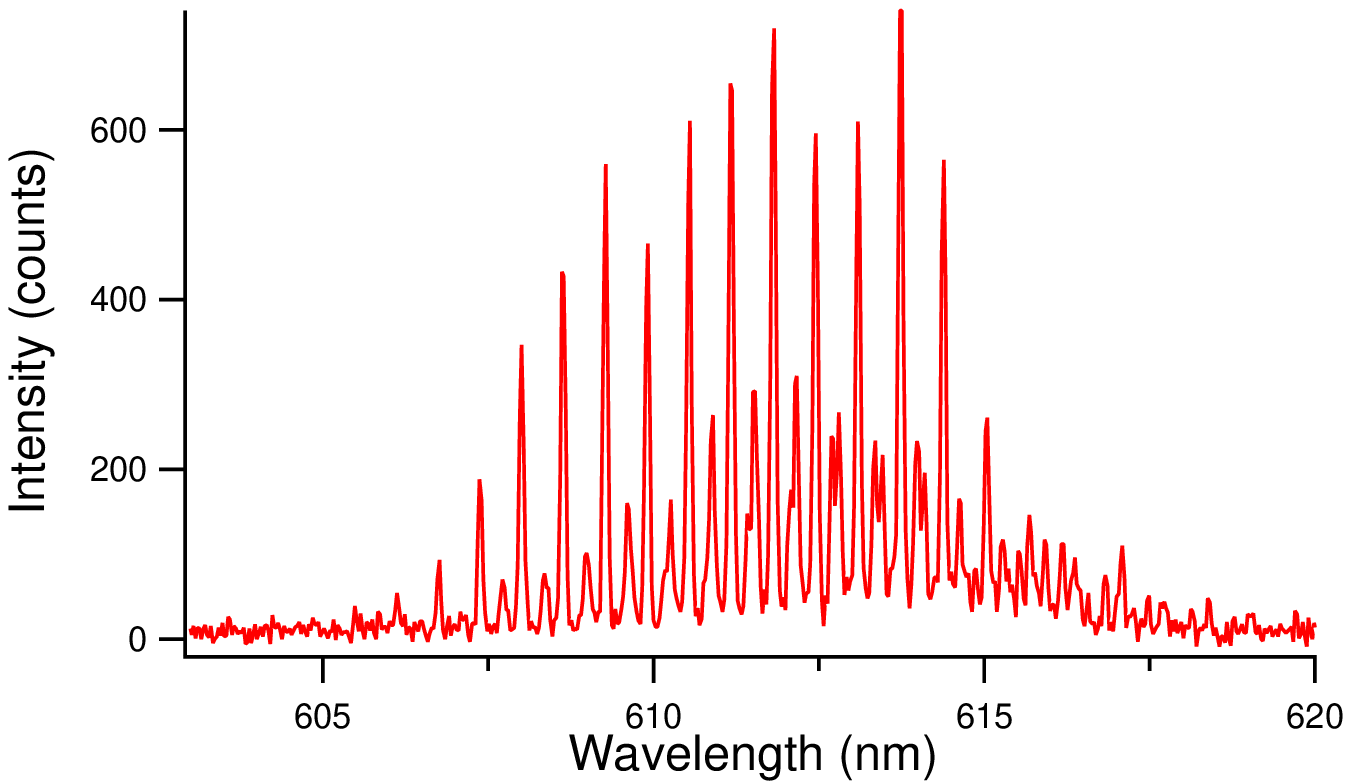}
\begin{center}(a)\end{center}
\end{minipage}\hfill
\begin{minipage}[t!]{.8\linewidth}
\includegraphics[width=1\linewidth]{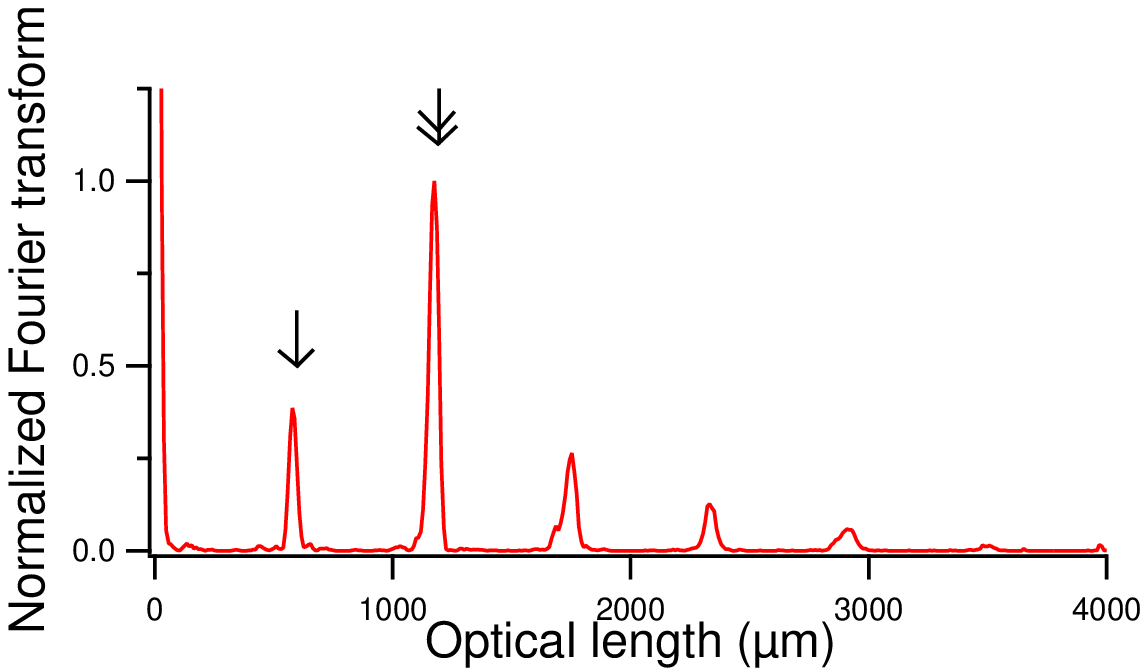}
\begin{center}(b)\end{center}
\end{minipage}\hfill
\caption{(a) Experimental spectrum from a pentagonal micro-laser
with $a=90\,\mu$m. (b) Fourier transform of the spectrum in (a). The
single (resp. double) arrow indicates the expected position of the
single (resp. double) pentagon periodic orbit.}
    \label{fig:penta-manips}
\end{figure}

A typical experimental spectrum from a pentagonal micro-laser is
plotted in Fig.~\ref{fig:penta-manips}a. Its Fourier transform
presented in Fig.~\ref{fig:penta-manips}b is mostly peaked at the
length of the double pentagon periodic orbit. The single pentagon is
visible as well, which can be directly noticed on the spectrum made
of two combs of different amplitudes. In \cite{lebental}, we
reported an experimental spectrum from a pentagonal micro-laser
where both combs had similar amplitudes and therefore its Fourier
transform did not present any peak at the length of the single
pentagon. The parameters which control the relative amplitudes of
the combs have not been identified yet, but it is clear that the
etching quality is a key point.

\section{Chaotic dielectric cavities}\label{forth}

Eventually we applied the same ideas to an archetypal chaotic
cavity, the Bunimovich stadium, which is made of a rectangle between
two half circles (see Fig.~\ref{fig:spectre-stade}b for notations),
and investigated various deformation defined by the parameter
$\rho=L/R$.

\subsection{Numerics}

\begin{figure}[!hbt]
  \begin{minipage}[r]{.99\linewidth}
    \includegraphics[width=1\linewidth]{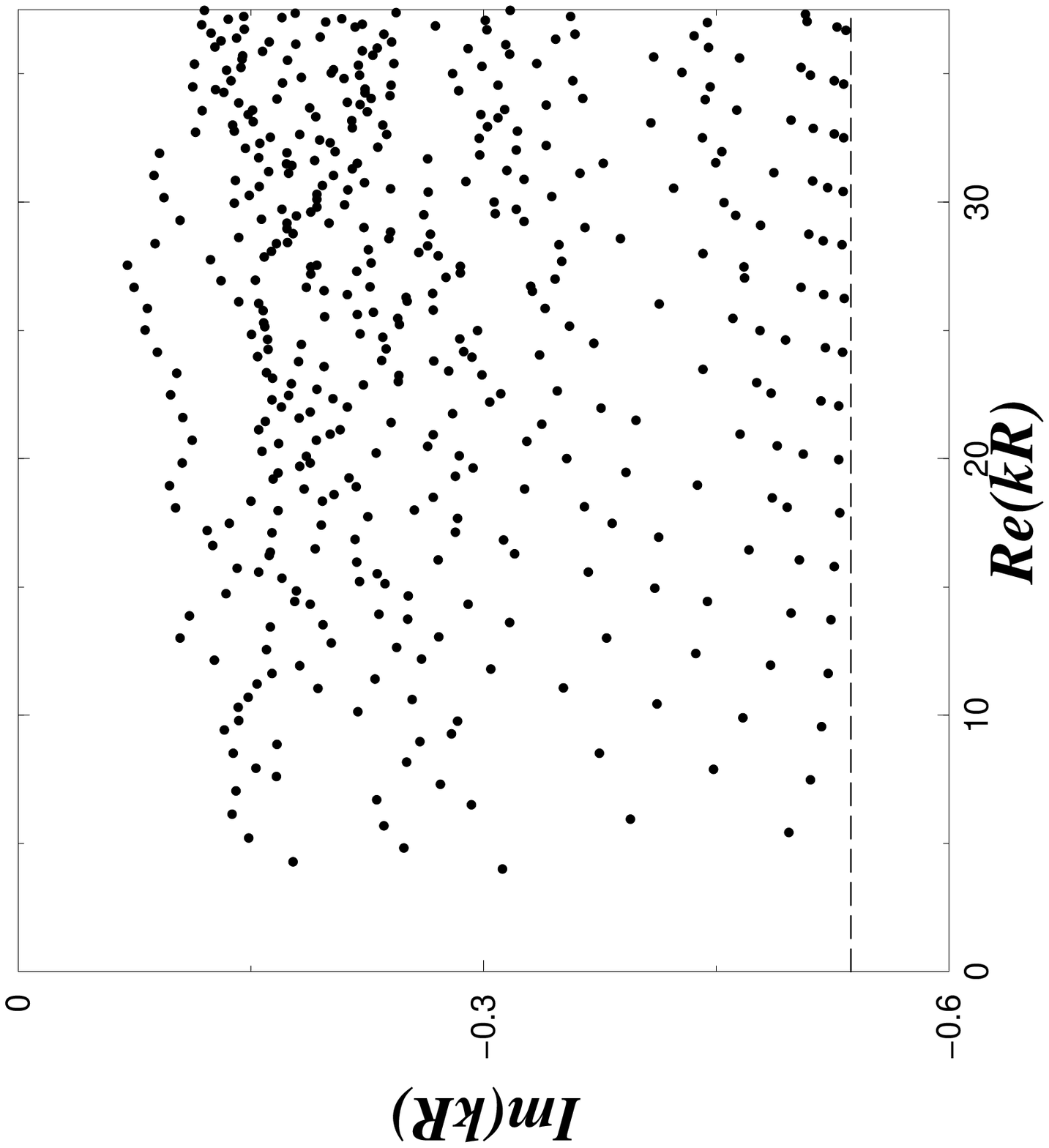}
\begin{center}(a)\end{center}
 \end{minipage}
\begin{minipage}[l]{0.4\linewidth}
    \includegraphics[width=1\linewidth]{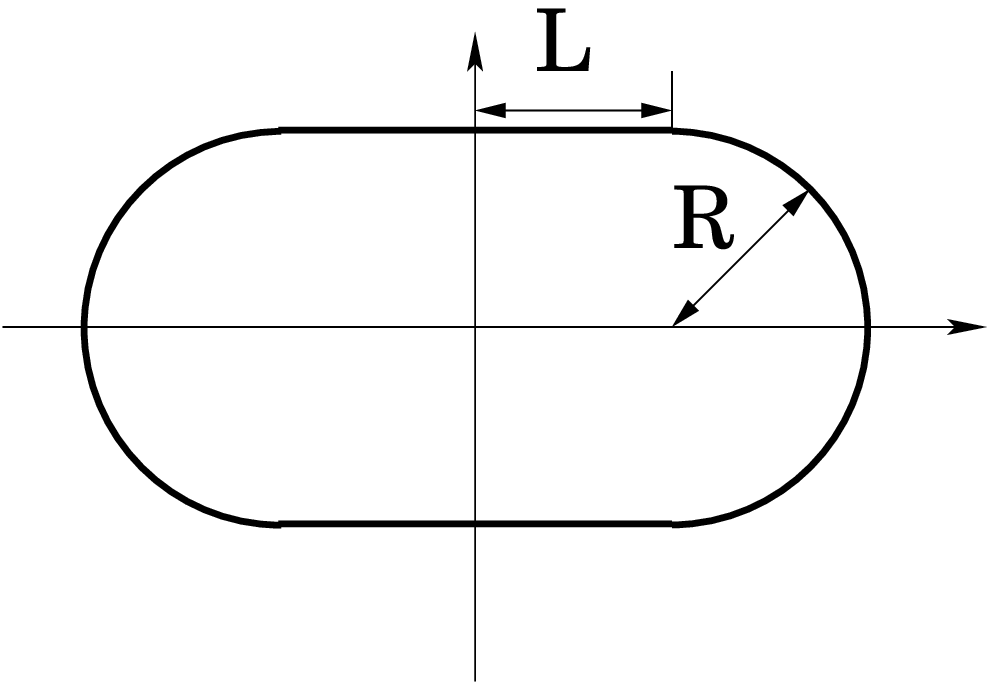}
\vspace{.9cm}
\begin{center}(b)\end{center}
  \end{minipage}
\begin{minipage}[l]{0.55\linewidth}
    \includegraphics[width=1\linewidth]{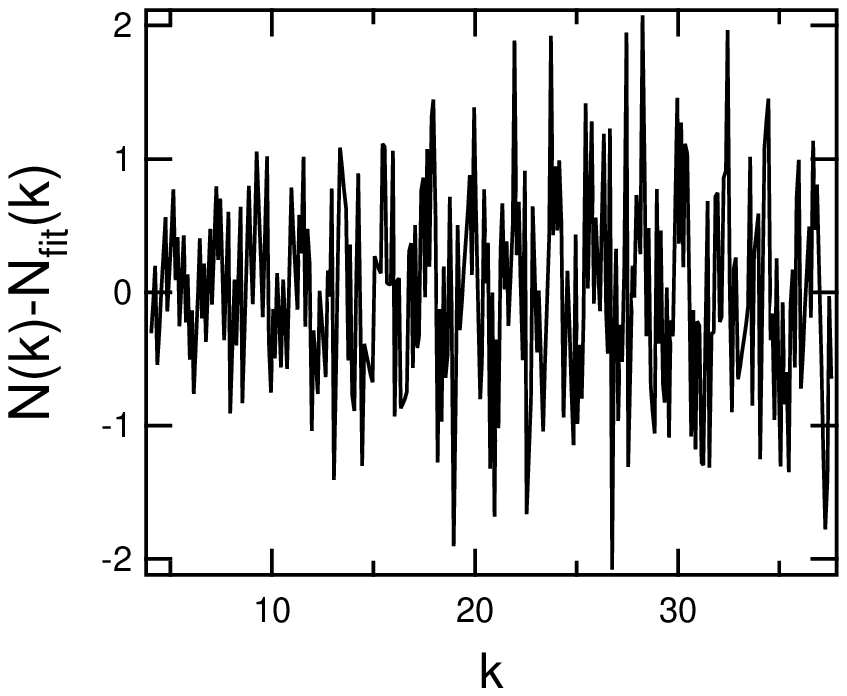}
\begin{center}(c)\end{center}
  \end{minipage}
\caption{(a) Resonance spectrum for the dielectric stadium with
$\rho=1$, $(--)$ symmetry class, and $n=1.5$. The position of the
horizontal dashed line is given by (\ref{eq:pertes-stade}). (b)
Notations for the Bunimovich stadium. (c) $N(k)-N_{fit}(k)$
calculated from the numerical simulations in (a) with $R=1$.}
\label{fig:spectre-stade}
\end{figure}

\begin{figure}
\begin{minipage}[l]{1\linewidth}
\includegraphics[width=.5\linewidth]{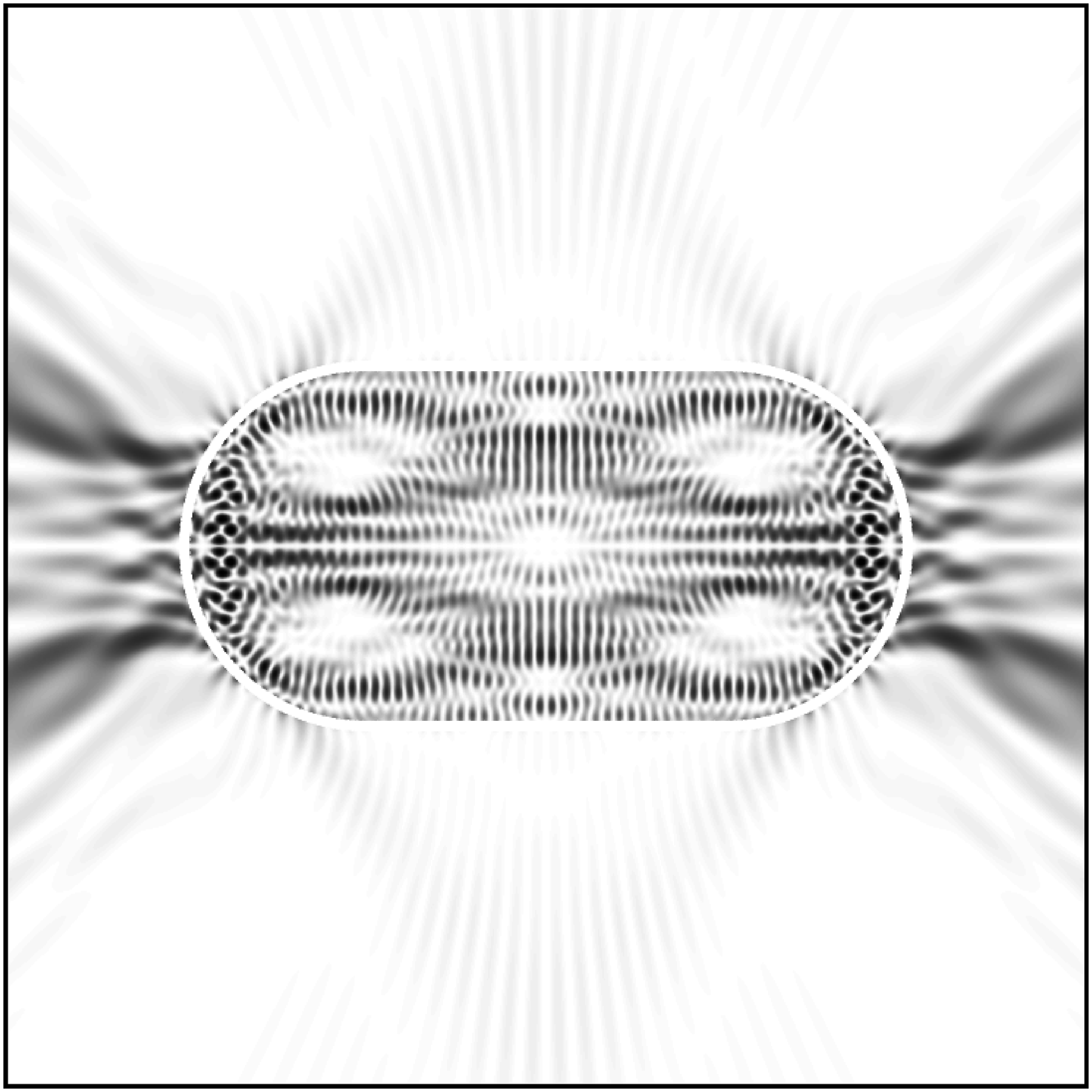}
\centering (a)
\end{minipage}
\begin{minipage}[r]{1\linewidth}
\includegraphics[width=.5\linewidth]{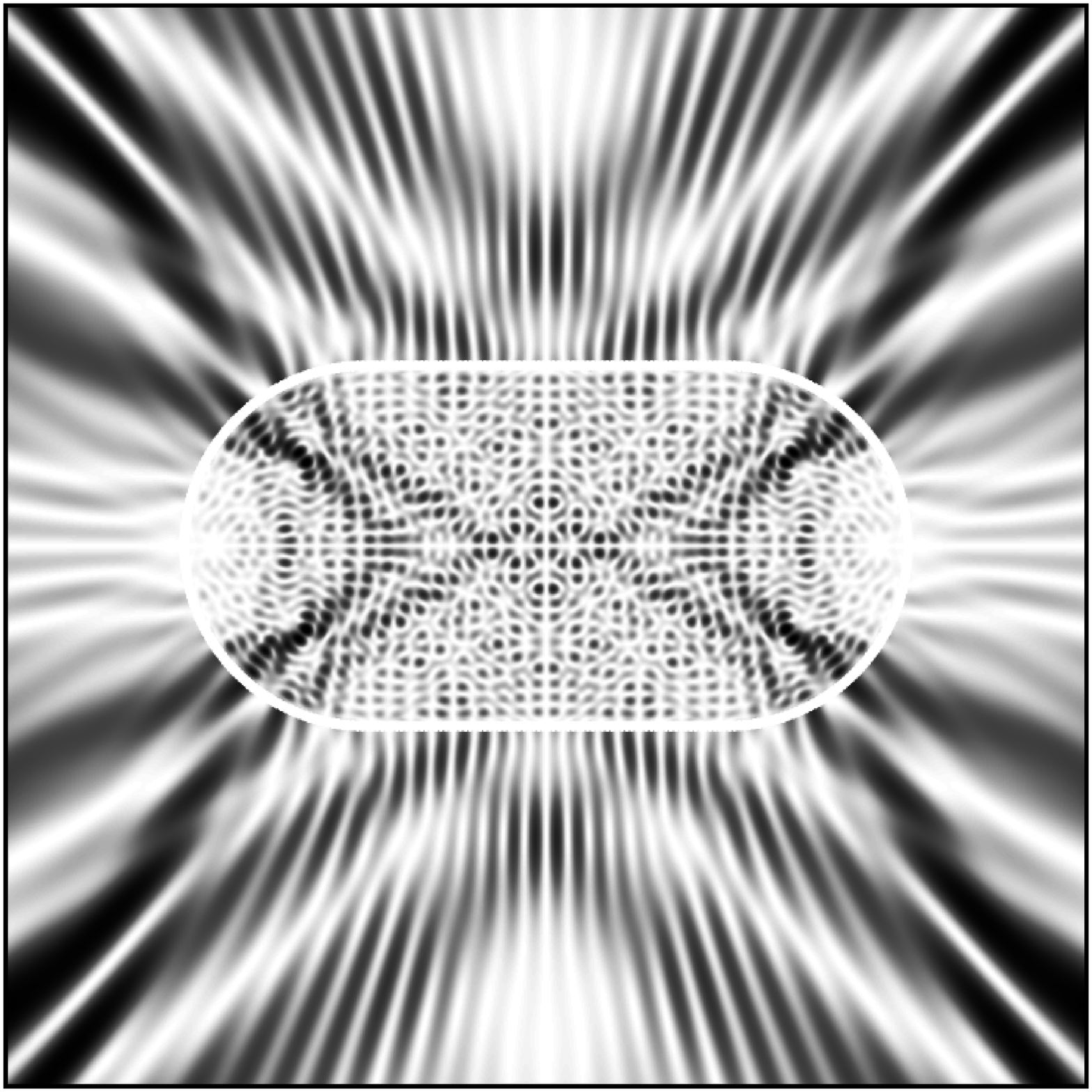}
\centering (b)
\end{minipage}
\begin{minipage}[r]{1\linewidth}
\includegraphics[width=.5\linewidth]{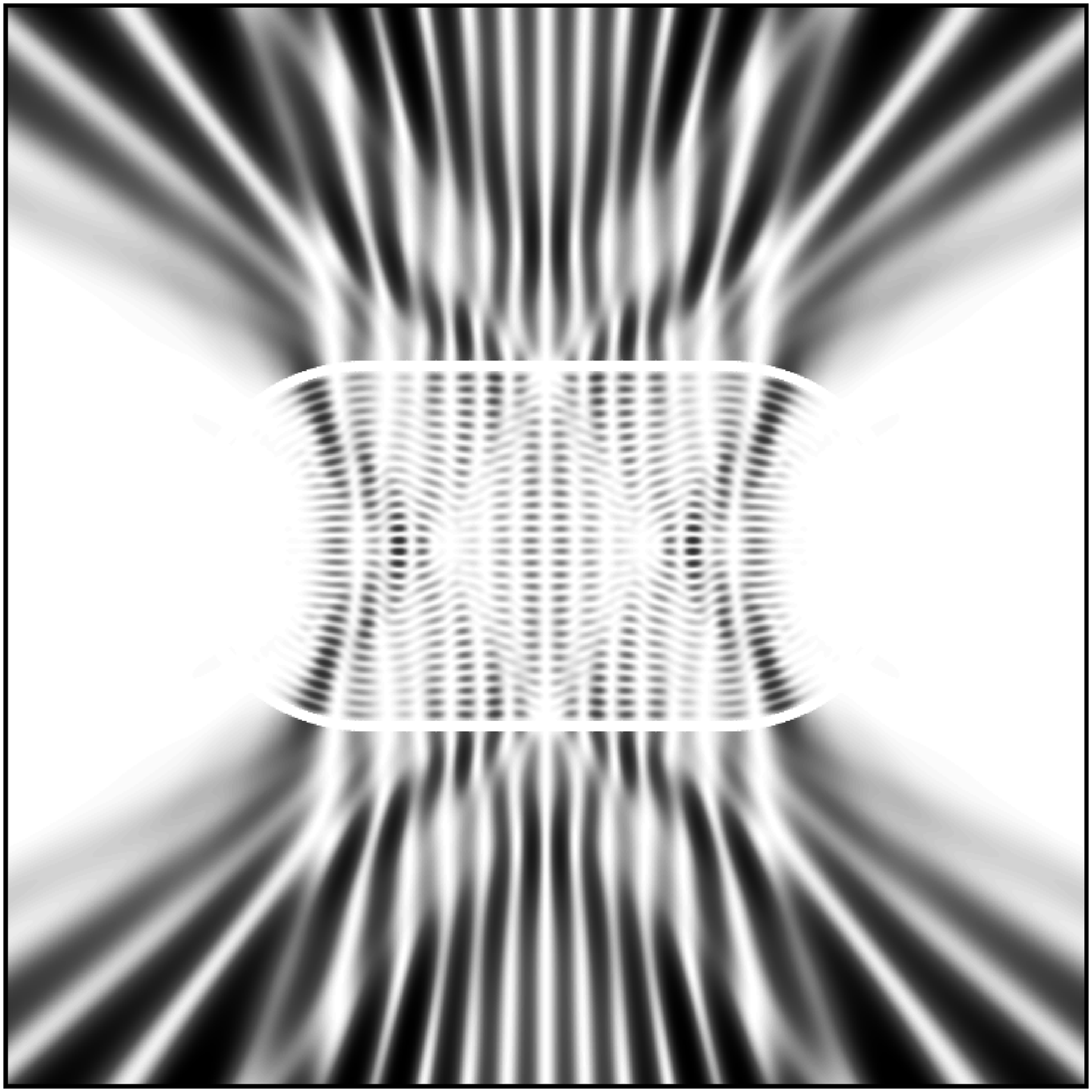}
\centering (c)
\end{minipage}
\caption{Wave functions of the dielectric stadium with $\rho=1$,
$(--)$ symmetry
  class, and $n=1.5$.
(a) $kR=34.48 -0.11\,\ic$.  (b) $kR=34.55 -0.31\,\ic$. (c) $kR=34.48
-0.45\,\ic$. Greyscale, black representing maximal values of
$|\psi|^2$.} \label{psi_stade}
\end{figure}

For simplicity we only consider the $(--)$ symmetry class, which
means that the associated wave functions vanish along both symmetry
axis of the stadium. The resonance spectrum for $\rho=1$ is shown in
Fig.~\ref{fig:spectre-stade}a. As for the other cavities, the
imaginary part of the resonances is bounded and its lower bound
$\gamma_{max}$ can be estimated from the refractive losses of the
periodic orbit which presents the highest losses, i.e. the
Fabry-Perot along the small axis (also called bouncing ball orbit):
\begin{equation}
  \label{eq:pertes-stade}
  \gamma_{max}= \frac{1}{2n}\ln\left(\frac{n-1}{n+1}\right)
\end{equation}
Again, some wave-functions are presented in Fig. \ref{psi_stade},
representative of different parts of the spectrum.

The counting function $N(k)$ was computed from the numerical data
shown in Fig.~\ref{fig:spectre-stade}a and the best fit gives:
\begin{equation}
  \label{weylstade}
  N_{fit}(k)=\frac{n^2}{4\pi}\left(\rho+\frac{\pi}{4}\right)
  (kR)^2-0.145 \,kR-4.042\ ,
\end{equation}
to be compared with the prediction (\ref{eq:smooth-lin}) for the linear term:
\begin{eqnarray}
\alpha^{th}&=& \frac{1}{4\pi}\left[\tilde{r}(n)\Big( \rho +\frac{\pi}{2}
    \Big)-n\,(2+\rho) \right]\Big |_{n=1.5,\rho=1} \nonumber\\
& \simeq &-.148 \ .
\label{linearstad}
\end{eqnarray}

\begin{figure}
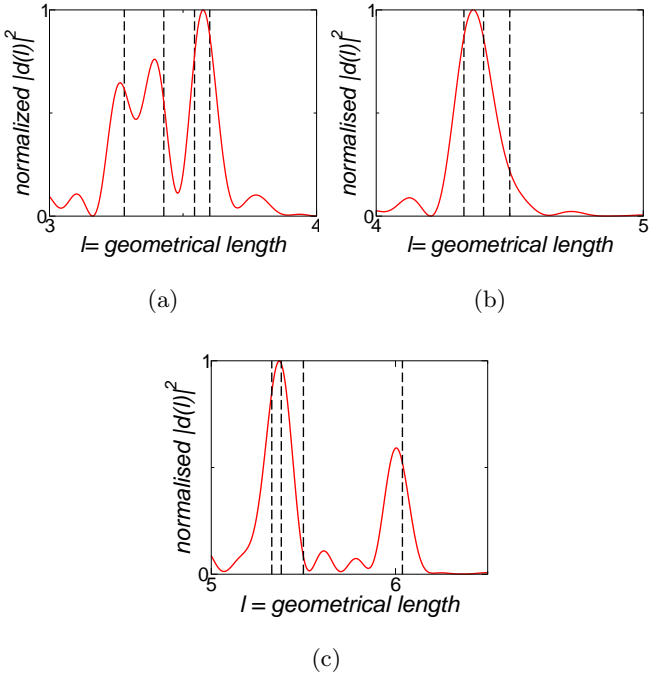

\begin{minipage}[l]{0.49\linewidth}
\includegraphics[width=1\linewidth]{fig50.eps}
\begin{center}(a)\end{center}
\end{minipage}
\vspace{0.5cm}
\begin{minipage}[r]{0.49\linewidth}
\includegraphics[width=1\linewidth]{fig51.eps}
\begin{center}(b)\end{center}
\end{minipage}
\begin{minipage}[r]{1\linewidth}
\includegraphics[width=.5\linewidth]{fig52.eps}
\begin{center}(c)\end{center}
\end{minipage}
\caption{Length densities calculated from numerical spectra for
dielectric stadiums with $(--)$ symmetry class and $n=1.5$. (a)
$\rho=0.3$. (b) $\rho=0.75$. (c) $\rho=1.25$. The vertical dotted
lines indicate the expected positions of periodic orbits (numbering
according to Fig.~\ref{tab:listorb}).} \label{fig:d_l_stade}
\end{figure}

\begin{figure}
\includegraphics[width=1\linewidth]{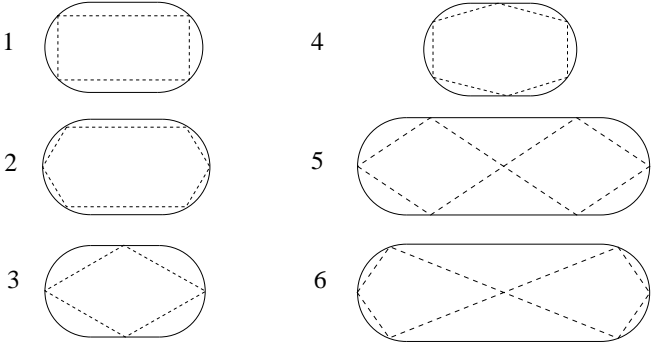}
  \caption{Some periodic orbits of the stadium.}
  \label{tab:listorb}
\end{figure}

\begin{figure}
\includegraphics[width=0.6\linewidth]{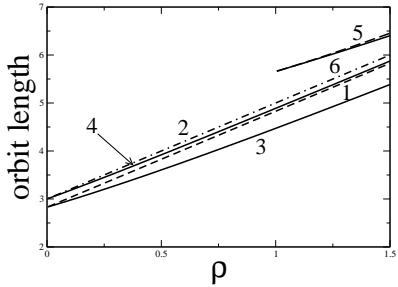}
  \caption{Geometrical length of the periodic orbits plotted in Fig.
  \ref{tab:listorb}. Numbering according to Fig. \ref{tab:listorb}.}
  \label{fig:longueur-orbites}
\end{figure}

\begin{figure}
\begin{minipage}[l]{.49\linewidth}
\centering
\includegraphics[width=1\linewidth]{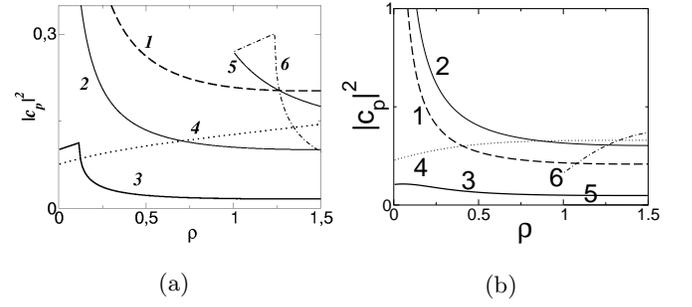}
\begin{center}(a)\end{center}
\end{minipage}
\begin{minipage}[r]{.49\linewidth}
\centering
\includegraphics[width=1\linewidth]{fig55.eps}
\begin{center}(b)\end{center}
\end{minipage}
\caption{Amplitude of the orbits listed in Tab.~\ref{tab:listorb}
for the $(--)$ symmetry class following formula (\ref{eq:cp-isolee})
(a) Semiclassical regime (i.e. no size effect taken into account).
(b) Idem with curvature correction calculated for $kR=25$.}
\label{fig:cp_stade}
\end{figure}

The oscillatory part of the trace formula is also checked, plotting
the length densities calculated from numerical spectra for several
shape ratios $\rho$. The curves presented in
Fig.~\ref{fig:d_l_stade} are peaked at different positions which
could be assigned to periodic orbits. To predict which periodic
orbits should mainly contribute to the length density, we calculated
their weighting coefficient from formula (\ref{eq:cp-isolee}). The considered orbits are drawn in
Fig.~\ref{tab:listorb}, their geometrical length is plotted in Fig.
\ref{fig:longueur-orbites}, and their coefficient (amplitude) versus
$\rho$ in Fig.~\ref{fig:cp_stade}a. Note
that orbits $5$ and $6$ obey geometrical constraints such as they do
not exist for $\rho < 1$.

\begin{figure}
\centering\includegraphics[width=0.9\linewidth]{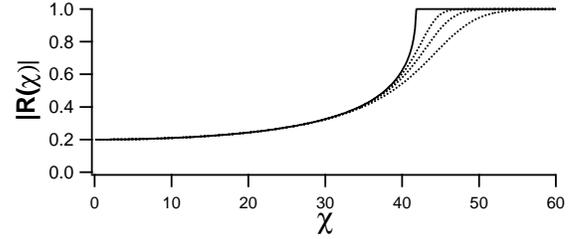}
\caption{Modulus of the TM reflection coefficient as a function of
the incidence angle (in degrees). Full line: Fresnel coefficient
  (\ref{eq:fresnel}) for a straight boundary. Dotted lines:
  (\ref{reflex_courb}) for $kR=100, 50, 25$.}
\label{fresnelcourb}
\end{figure}

The length densities are calculated from the numerical spectra on a
finite number of resonances (finite range of Re($kR$)), thus some
finite size effects do play a role and must be taken into account
evaluating the amplitudes of the periodic orbits. One of the main
effect here comes from the curvature correction in the
reflection coefficient. If the dielectric boundary is curved enough
compared to the wavelength then there is a quite important
correction to the standard Fresnel coefficients \cite{marthenn}. We
use the following formula to take into account the curvature
correction:
\begin{equation}
  \label{reflex_courb}
  R^{(c)}_{TM}=\frac{\sqrt{n^2-\dfrac{m^2}{x^2}} +\ic
    \dfrac{H^{(1)\prime}_m}{H^{(1)}_m}(x)}{
    \sqrt{n^2-\dfrac{m^2}{x^2}} -\ic
    \dfrac{H^{(1)\prime}_m}{H^{(1)}_m}(x)}
\end{equation}
where $x=kR$ and $m=nx\cos\chi$. The Fresnel coefficient for a
straight interface (\ref{eq:fresnel}) is recovered for large
$kR$, see Fig.~\ref{fresnelcourb}. In Fig.~\ref{fig:cp_stade}b,
the amplitudes of the periodic orbits are plotted taking into
account this curvature correction. Notice the important differences with
Fig.~\ref{fig:cp_stade}a.

Using Fig.~\ref{fig:cp_stade}b we can give a quantitative estimate
of the periodic orbits which mostly contribute to the length
density. For $\rho=0.3$ (Fig.~\ref{fig:d_l_stade}a), the length
density is peaked around the orbits 3 ($l\simeq 3.28$) and 1
($l\simeq 3.43$), and orbits 2 and 4 with respective length
$l\simeq3.6$ and $l\simeq 3.54$ cannot be separated. For $\rho=0.75$
(Fig.~\ref{fig:d_l_stade}b), the orbits $1$ and $4$ with respective
length $l\simeq 4.33$ and $l\simeq 4.40$ interfere. The line at
$l\simeq4.5$ stands for orbit $2$. Eventually for $\rho=1.25$ (Fig.
\ref{fig:d_l_stade}c), the two orbits 1 and 4 ($l\simeq 5.33$ and
$l\simeq 5.38$) interfere. Again we drew a line for orbit 2 at
$l\simeq5.5$. A peak can also be seen for orbit 6 ($l\simeq 6.04$).
From these examples it appears that the agreement between theory and
numerics is qualitatively good.

\subsection{Experiments}

\begin{figure}[!hbt]
\includegraphics[width=0.9\linewidth]{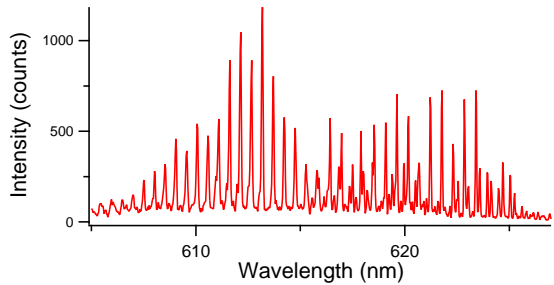}
\caption{Experimental spectrum of a stadium micro-laser with
$\rho=0.75$ and $R=50\,\mu m$. Its Fourier transform is plotted in
Fig. \ref{fig:manips-stade-tf}b.} \label{fig:manips-stade-spectre}
\end{figure}

\begin{figure}[!hbt]
\includegraphics[width=0.9\linewidth]{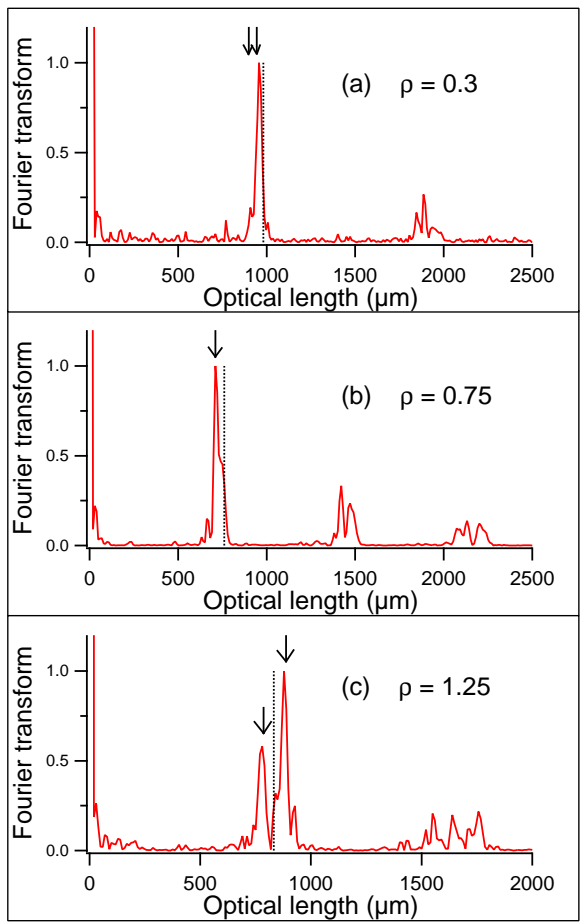}
\caption{Normalized Fourier transform of experimental spectra from
stadium micro-lasers with (a) $R=80\,\mu m$, (b) $R=50\,\mu m$, and
(c) $R=45\,\mu m$. The vertical line corresponds to the position of
the perimeter. The arrows indicate the expected positions of some
orbits, for comparison: (a) orbits 1 and 2, (b) orbit 1, and (c)
orbits 1 and 5 (numbering according to Fig. \ref{tab:listorb}).}
\label{fig:manips-stade-tf}
\end{figure}

Spectra were recorded for the shape ratios used in numerical
simulations and for various $R$ (see \cite{djellali2} for
$\rho=0.5$). A typical experimental spectrum from an organic
micro-stadium is plotted in Fig. \ref{fig:manips-stade-spectre} and
typical Fourier transforms in Fig. \ref{fig:manips-stade-tf}. They
look similar to those of the shapes studied in the previous sections. However the main
difficulty to face when studying stadiums is the large number of
orbits with close lengths. Would it be numerically or
experimentally, it is thus difficult to assign peaks in the length
density. Then, to check formula (\ref{eq:cp-isolee}) against
experiments, it was decided to compare the positions of the peaks to
the length of the perimeter. Actually some crossing orbits like
orbits 5 and 6 are longer than the perimeter and, due to geometrical
constraints, do not exist for $\rho<1$. Moreover according to Fig.
\ref{fig:cp_stade}a, which corresponds to the semi-classical limit
and so to experimental conditions, their amplitudes
(\ref{eq:cp-isolee}) are the highest when they appear. So we expect
Fourier transforms peaked at positions shorter than the perimeter
for $\rho<1$ and longer for $\rho\gtrsim 1$, and this is evidenced
in Fig. \ref{fig:manips-stade-tf}.

\section{Conclusion}

In this paper we have shown numerical and experimental results
concerning the trace formula for dielectric cavities. For convex
cavities and TM polarization the resonance spectrum can be divided
into two subsets. One of them, the Feschbach (inner) resonances
which are relevant for experiments, is statistically well described
by classical features: the periodic orbit with the shortest lifetime
for the lower bound of the wave number imaginary parts, the Weyl's
law for the counting function, and the weighting coefficients of
periodic orbits for the length density. The formul\ae{} we derived,
based on standard expressions used in quantum chaos and adapted to
dielectric resonators, give an accurate description of these
spectral properties.

Those formul\ae{} have been checked for various resonator shapes.
For "regular shape" (i.e. the analogous billiard problem is
separable) and for small index of refraction, the resonance spectrum
presents a branch structure as if the dielectric problem was
separable. When the corresponding billiard problem is not
integrable, the usual Weyl's law still occurs. The oscillating part
can be also explained by taking into account  the shortest periodic
orbits. In the chaotic case the correspondence is however more
difficult to claim quantitatively as finite size effects play a
quite important role because of periodic orbits with close lengths.

The main  result of the paper is the demonstration that dielectric
cavities widely used in optics and photonics can be well
described using generalizations of techniques from quantum chaos.

This study raises many open problems and we would like to mention
some of them. The formul\ae{} checked here gave accurate predictions
for TM polarization.  For TE polarization, due to the existence of
the Brewster angle where the reflection coefficient vanishes
\cite{dettmann}, the situation is less clear and requires further
investigations. The next step should be  to treat carefully the
diffraction on dielectric wedges, which is still an open problem
\cite{burge}. A related question is to improve the accuracy of the
standard effective index theory, since the separation into TE and TM
polarizations is precisely based on this 2D approximation.
Experimental data (i.e. real 3D systems) indeed revealed departures
from this model \cite{gozhyk,bittner}. These questions are related
to the wave functions (see e.g. \cite{lebental}) and far field
patterns, which are also of great interest, especially for
applications.

\section*{Acknowledgments}

The authors are grateful to D. Bouche, S. Lozenko, C. Lafargue, and
J. Lautru for fruitful discussions and technical support.

\appendix

\section{Weyl's law}\label{sec:app-weyl}

The Appendix deals with the derivation of formula (\ref{eq:smooth})
using an alternative method than in \cite{bogomolny}. Start from the
definition of the Green function $G$:
\begin{equation}
  \label{def_G}
  (\Delta_{\vec{x}} + n(\vec{x}\,)^2 k^2)\,G(\vec{x},\vec{y}\,)=\delta(\vec{x}-\vec{y}\,)
\end{equation}
where $n(\vec{x}\,)$ is equal to $n$ (resp. $1$) when $\vec{x}$ is
inside (resp. outside) the dielectric cavity. Moreover, for TM
modes, $G(\vec{x},\vec{y}\,)$ and its normal derivative are continuous
along the boundary of the domain. Taking the trace of it gives the
density of states through the Krein formula (see \cite{bogomolny} and references therein):
\begin{eqnarray}
  \label{krein}
 & & d_{all}(E)-d_0(E)=\\
&-&\frac{1}{\pi}\int \im \left[ n(\vec{x}\,)^2
G(\vec{x},\vec{x}\,) -G_0(\vec{x},\vec{x}\,)\right]\ud\vec{x}
\nonumber
\end{eqnarray}
where the integral runs over the whole 2D plane. $d_0(E)$ is the
density of states of the free space and $G_0(\vec{x},\vec{y}\,)$
stands for the Green function of a free particle in the plane:
\begin{eqnarray}
  \label{free_G}
 && G_0(\vec{x},\vec{y}\,)=\frac{1}{4\ic}\,H_0^{(1)}(k\left| \vec{x}-\vec{y}\, \right|)\\
  &&= \frac{1}{4\pi\ic}\int_{-\infty}^{+\infty}
\frac{e^{\ic p (x_1-y_1)}}{\sqrt{ k^2-p^2}}
  e^{\ic\sqrt{k^2-p^2}\,|x_2-y_2|}\ud p
\label{eq:H0-developpe}
\end{eqnarray}
with $\vec{x}=(x_1,x_2)$ and $\vec{y}=(y_1,y_2)$. $H_0^{(1)}(z)$ is
the Hankel function of the first kind and
Eq.~(\ref{eq:H0-developpe}) implicitly assumes that $k$ has a small
positive imaginary part. It is worth noting that $d_{all}(E)$ is the
Krein spectral shift function which is different from the spectral
density $d(E)$ discussed in Eq.~(\ref{eq:densite-generale}). It will
be explained how to get it at the end of this Section.

\subsection{Derivation of the first two terms}

\begin{figure}
\includegraphics[width=1\linewidth]{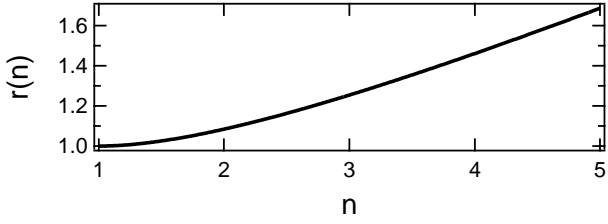}
\caption{Plot of $\tilde{r}(n)$.} \label{fig:rtilde}
\end{figure}

The leading term of the Weyl's law is obtained when substituting
$G_0$ to $G$ in Eq. (\ref{krein}) and using (\ref{free_G}):
\begin{equation}
  \label{weyl1}
  d_{all}(E)-d_0(E)\simeq (n^2-1)\frac{{\cal A}}{4\pi}
\end{equation}
where ${\cal A}$ is the area of the domain filled with the
dielectric material.

The first remaining term in (\ref{eq:smooth}) comes from the
presence of the boundary. Thus it is first necessary to solve the
elementary problem of a plane wave reflecting on an infinite
straight dielectric boundary, which can be derived through standard
methods. Then, using expression (\ref{eq:H0-developpe}), the Green
function for both $\vec{x}$ and $\vec{y}$ inside the dielectric can
be written:
\begin{eqnarray}
  G(\vec{x},\vec{y}\,)&=&\frac{1}{4\pi\ic}\int_{-\infty}^{+\infty}
\frac{e^{\ic p (x_1-y_1)}}{\sqrt{n^2 k^2-p^2}}[
  e^{\ic\sqrt{n^2 k^2-p^2}\ |x_2-y_2|}
  \nonumber\\ &+&R(p) \,e^{\ic \sqrt{n^2 k^2-p^2}\,
    |x_2+y_2|} ] \,\ud p\label{G_plane_in}
\end{eqnarray}
where $p$ is the tangential component of the momentum and
\begin{equation}
  \label{fresnel2}
  R(p)=\frac{\sqrt{n^2 k^2 -p^2}-\sqrt{k^2 -p^2}}{\sqrt{n^2 k^2
    -p^2}+ \sqrt{k^2 -p^2}}\ .
\end{equation}
Similarly one gets the Green function when the arguments are outside
the dielectric:
\begin{eqnarray}
  G(\vec{x},\vec{y}\,)&=& \frac{1}{4\pi\ic}\int_{-\infty}^{+\infty}
\frac{e^{\ic p (x_1-y_1)}}{\sqrt{ k^2-p^2}}[
  e^{\ic\sqrt{k^2-p^2}\ |x_2-y_2|}
  \nonumber\\&-&R(p)\, e^{\ic \sqrt{k^2-p^2}\
    |x_2+y_2|} ]\,\ud p \ .
\label{G_plane_out}
\end{eqnarray}
As usual the trace of $G$ is computed using local coordinates. The
surface term (\ref{weyl1}) is recovered from the first terms of
(\ref{G_plane_in}) and (\ref{G_plane_out}), so we focus now on their
second terms only. The integration along the boundary gives the
length factor ${\cal L}$. For the transverse coordinate, the
boundary is approximated locally by its tangent plane, and then
(\ref{G_plane_in}) and (\ref{G_plane_out}) are used. After the
convenient Wick rotation $p\to-itk$, the boundary contribution from
inside is:
\begin{equation}
  \label{bound_in}
   \alpha^{(in)}=\dfrac{n^2}{8\pi^2 k}\displaystyle\int_{-\infty}^{+\infty}
\frac{R(-\ic tk)}{n^2+t^2}\ud t\ .
\end{equation}
Similarly from (\ref{G_plane_out}) the boundary contribution is:
\begin{equation}
  \alpha^{(out)}=-\dfrac{1}{8\pi^2 k}\displaystyle\int_{-\infty}^{+\infty}
\frac{R(-\ic tk)}{1+t^2}\ud t\ . \label{bound_out}
\end{equation}
Putting together (\ref{bound_in}) and (\ref{bound_out}) back to
(\ref{krein}), one gets the first two terms of the Weyl expansion:
\begin{equation}
  \label{weyl2}
  d_{all}(E)-d_0(E)\simeq (n^2-1)\frac{{\cal
      A}}{4\pi}+\frac{(\tilde{r}(n)-1){\cal L}}{8\pi k}\ ,
\end{equation}
where, noting $R(t)$ instead of $R(-\ic t k)$:
\begin{equation}
 \tilde{r}(n)=1+\frac{n^2}{\pi}\int_{-\infty}^{+\infty}\frac{\mathrm{d}t}{t^2+n^2}R(t)-
 \frac{1}{\pi}\int_{-\infty}^{+\infty}\frac{\mathrm{d}t}{t^2+1}R(t)\ .
 \label{new_weyl}
\end{equation}
$\tilde{r}(n)$ is plotted in Fig. \ref{fig:rtilde}.

\subsection{From the density of states to the count of Feschbach resonances}

As mentioned above, the quantity entering the Krein formula is not
exactly the spectral density (\ref{eq:densite-generale}), but they
can be related heuristically. The spectral shift function,
$d_{all}(E)$ in (\ref{krein}), is related with the determinant of
the full $S$-matrix for the scattering on a cavity, while $d(E)$ in
(\ref{eq:densite-generale})  is the spectral density of Feschbach
(inner) resonances which are poles of this $S$-matrix.  In addition
to these poles, the determinant of the $S$-matrix may have poles
associated with shape resonances (which we do not take into account)
and then an additional phase factor $d_s(E)$:
\begin{equation}
d_{all}(E)=d(E)+d_s(E).
\end{equation}
It is natural to assume (and can be checked for dielectric disk)
that for all outside structures the corresponding wave functions are
almost zero inside the cavity and on its boundary. Therefore the
$S$-matrix phase, $d_s(E)$, associated with such functions in the
leading order is the same as for the outside scattering on the same
cavity but with the Dirichlet boundary conditions. The Weyl
expansion for such scattering is known (see
\cite{petkovpopov,robert2} and references therein):
\begin{equation}
  \label{weyl_out}
  d_s(E)-d_0(E)\simeq -\frac{{\cal A}}{4\pi}-\frac{{\cal L}}{8\pi k}\ .
\end{equation}
Subtracting (\ref{weyl_out}) to (\ref{weyl2}) gives the desired
result for the spectral density, taking into account only Feschbach
resonances:
\begin{equation}
  \label{weyl_final}
  d(k)=2k d(E)\simeq n^2\frac{{\cal A}k}{2\pi}+\frac{\tilde{r}(n)}{4\pi}\,{\cal L}\ .
\end{equation}
Then Eq.~(\ref{eq:smooth}) is recovered by integration.


\begin{thebibliography}{99}

\bibitem{vahala} K. J. Vahala, 
  Nature \textbf{424}, 839 (2003)

\bibitem{matsko} A.B. Matsko, \emph{Practical applications of microresonators
    in optics and photonics}, CRC Press (2009)

\bibitem{bogomolny} E. Bogomolny, R. Dubertrand, C. Schmit, Phys. Rev. E,
  \textbf{78}, 056202 (2008)

\bibitem{laxphillips} P. Lax, R. S. Phillips, \emph{Scattering
    Theory}, Springer NY (1963)

\bibitem{petkovpopov} V. Petkov, G. Popov, Ann. Inst. Fourier
  \textbf{32}, 111 (1982)

\bibitem{uzy} U. Smilansky and I. Ussishkin, J. Phys. A: Math. Gen. \textbf{29}, 2587 (1996)

\bibitem{robert1} D. Robert, \emph{Partial differential equations and
    mathematical physics}, Birkh\"auser Boston (1996)

\bibitem{robert2} D. Robert, Helv. Phys. Acta \textbf{71}, 44 (1998)

\bibitem{popovvodev} G. Popov, G. Vodev, Comm. in
  Math. Phys. \textbf{207}, 411 (1999)

\bibitem{lebental} M. Lebental, N. Djellali, C. Arnaud, J.-S. Lauret,
  J. Zyss, R. Dubertrand, C. Schmit, E. Bogomolny, Phys. Rev. A,
  \textbf{76}, 023830 (2007).

\bibitem{lebental-matsko}
M. Lebental, E. Bogomolny, and J. Zyss, in \emph{Practical
applications of microresonators in optics and photonics}, A. Matsko,
CRC Press (Boca Raton, 2009).

\bibitem{bittner_2} S. Bittner, B. Dietz, M. Miski-Oglu, Oria Iriarte,
  A. Richter, F. Sch\"afer, Phys. Rev. A \textbf{80}, 023825 (2009).


\bibitem{sjostrand} J. Sj\"ostrand and M. Zworski, Acta Math. \textbf{183}, 191
(1999).

\bibitem{gutzwiller} M. Gutzwiller, \emph{Chaos in classical and quantum
  mechanics}, Springer Berlin (1990).

\bibitem{bb} R. Balian, C. Bloch, Ann. Phys. \textbf{60},
  401 (1970);  Ann. Phys. \textbf{64}, 271 (1971);
  Ann. Phys. \textbf{69}, 76 (1972).

\bibitem{chang}
R. K. Chang et A. J. Campillo eds., \emph{Optical processes in
microcavities}, World Scientific (1996).

\bibitem{micro-ondes}D. Kajfez and P. Guillon, \emph{Dielectric
resonators}, Norwood MA (1986).

\bibitem{preu}
S. Preu, H. G. L. Schwefel, S. Malzer, G. H. Dhler, L. J. Wang, M.
Hanson, J. D. Zimmerman, and A. C. Gossard, Optics Express, {\bf
16,} 7336 (2008).

\bibitem{gozhyk}
I. Gozhyk et al., to be published.

\bibitem{djellali}
S. Lozenko, N. Djellali, J. Lautru, I. Gozhyk, D. Bouche, M. Lebental,
C. Ulysse, and J. Zyss. Submitted.

\bibitem{lebental2}
M. Lebental, J.-S. Lauret, J. Zyss, C. Schmit, and E. Bogomolny,
 \pra {\bf 75,} 033806 (2007).

\bibitem{dubertrand} R. Dubertrand, E. Bogomolny, N. Djellali, M. Lebental,
  C. Schmit,  Phys. Rev. A \textbf{77}, 013804 (2008).

\bibitem{Eugene_square} E. Bogomolny et al, to be published


\bibitem{pseudoint} P. J. Richens \& M. V. Berry,
  Physica D \textbf{2}, 495 (1981)
\bibitem{barbara} B. Dietz, U. Smilansky, 
  Physica D \textbf{86}, 34 (1995)

\bibitem{bittner} S. Bittner, E. Bogomolny, B. Dietz, M. Miski-Oglu,
P. Iriarte, A. Richter, and F. Sch\"afer, \pre {\bf 81,} 066215
(2010).

\bibitem{bill_ell} E. Mathieu, J. Math. Pures et Appl. \textbf{13} (in
  French), 137 (1868)

\bibitem{olivier} E. Bogomolny, O. Giraud, C. Schmit, Comm. in
  Math. Phys. \textbf{222}, 327 (2001)

%

\bibitem{marthenn} M. Hentschel, H. Schomerus, Phys. Rev. E
  \textbf{65}, 045603 (2002).

\bibitem{djellali2} N. Djellali, I. Gozhyk, D. Owens, S. Lozenko,
M. Lebental, J. Lautru, C. Ulysse, B. Kippelen, and J. Zyss, \apl
{\bf 95,} 101108 (2009).

\bibitem{dettmann} C. P. Dettmann, G. V. Morozov, M. Sieber,
 H. Waalkens,
EPL \textbf{87}, 34003 (2009).

\bibitem{burge} R. Burge, X.-C. Yuan, B. Carroll, N. Fisher, T. Hall,
G. Lester, N. Taket, and C. Oliver, IEEE Transactions on antennas
and propagation, {\bf 47,} 1515 (1999).







\end{thebibliography}
\end{document}